\documentclass[11pt,twoside]{article}    

\usepackage{amssymb}
\usepackage{latexsym} 
\usepackage{mathptm}
\usepackage{epsfig}
\usepackage[dvips]{color}
\usepackage{epsf}

\date{} 
\begin{document}

\markboth{SABINE HOSSENFELDER}{WHAT BLACK HOLES CAN TEACH US}

\title{WHAT BLACK HOLES CAN TEACH US}
\author{Sabine Hossenfelder\thanks{sabine@physics.arizona.edu}\\
\small{Department of Physics}\\ 
\small{University of Arizona}\\
\small{1118 East 4th Street}\\ 
\small{Tucson, AZ 85721, USA}}

\date{}
\maketitle

\begin{abstract}Black holes merge together different field of physics. From General Relativity over 
thermodynamics and
quantum field theory, they do now also reach into the regime of particle and collider physics. 
In the presence of additional
compactified dimensions, it would be possible to produce tiny black holes at future colliders. 
We would be able to
test Planck scale physics and the onset of quantum gravity. The understanding of black hole physics
is a key knowledge to the phenomenology of these new effects beyond the Standard Model.

This article gives a brief introduction into the main issues and is addressed to a non-expert audience. 
\end{abstract}

\section{Introduction}

Gravity is different. Physicists have successfully described three of the known particle 
interactions in the
 Standard Model of particle physics. The electromagnetic interaction, which governs 
the motion of electrically charged particles, the weak interaction, which is e.g. responsible 
for the beta decay, and the strong interaction, which
bounds together the nuclei that we are made of. These three forces can be described as quantum 
field theories whose
mathematical traps and physical details are thought to be well known; if not understood then 
they are at least familiar.

Only gravity refuses to fit in the common treatment. That gravity is somewhat special can 
be experienced in daily
life. It is the only long range interaction and thus governs the dynamics of macroscopic 
objects, such as apples, planets 
or galaxies. Gravity is by far weaker than the other interactions. The strength of the 
electromagnetic force between two electrons is $\approx 10^{40}$ larger than the gravitational 
one! The only reason why we do observe it at all is
that the gravitational attraction can not be neutralized. The charge of gravity, 
the mass, is always positive. But for all purposes of particle physics at small distances, 
the other interactions dominate and gravity can, and is, safely neglected.

This article is organized as follows. In the next section we will introduce the Planck scale
and motivate its importance. Section three then reviews briefly some of the main features of 
black holes, including the evaporation characteristics. In Section four we will summarize 
the effective models with extra space time dimensions and examine some of the predicted
observables. Section five then deals with black holes in these models with extra dimensions, the
possibility of their production in the laboratory, their properties and observables. 
Section six investigates the formation of stable black hole remnants.
In Section \ref{MinimalLength} we connect black hole physic to the issue of a minimal length scale.
Section \ref{Dangerous} is dedicated to the recurring question whether the earth produced black
holes are dangerous. We conclude with a brief summary in Section \ref{Summary}.

Unless otherwise indicated, 
we assume natural units $\hbar=c=1$ which leads to $m_{\rm p}=1/l_{\rm p}$. The metric has
signature $(-1,1,1,...)$.

\section{The Planck Scale}

For the description of particle physics in the laboratory, interactions of the
Standard Model dominate and gravity can be safely neglected.
But things will change dramatically if we go to extreme regimes. Let us consider a test particle that is
accelerated to a very high energy $E$. As we know from quantum mechanics, such a particle can, and
must be, associated with a Compton wave length $\lambda=1/E$. On the other hand, 
Einstein's Theory
of General Relativity tells us that every kind of energy will cause space time to curve. 
This is described
by Einstein's Field Equations:
\begin{eqnarray}
R_{\mu \nu} - \frac{1}{2} g_{\mu \nu} R = - 8 \pi \frac{1}{m_{\rm p}^2}T_{\mu \nu} \quad. \label{EFG}
\end{eqnarray}
The sources of gravity are everything which carries energy and momentum. These quantities are described in the
energy momentum tensor $T_{\mu \nu}$ on the right hand side. The sources cause a curvature of spacetime which is
described by the metric $g_{\mu \nu}$, and some of its second derivatives in the curvature 
tensor\footnote{See Appendix A, Eq.(\ref{R1}), (\ref{R2}) and (\ref{R3}).} $R$. 
This in turn will then modify the motion of the sources in the very space time, giving a self consistent
time evolution of the system.

The equation of motions for a pointlike particle is the Geodesic equation:
\begin{eqnarray} \label{Geo}
\frac{{\mathrm{d}}^2x^{\mu}}{{\mathrm{d}}\tau^2}+
\Gamma^{\mu}_{\alpha \beta}\frac{{\mathrm{d}}x^{\alpha}}{{\mathrm{d}}\tau}
\frac{{\mathrm{d}}x^{\beta}}{{\mathrm{d}}\tau} =0 \quad ,
\end{eqnarray}
where the Christoffel-symbols  $\Gamma$ are defined in Eq. (\ref{Christoffel}) and $\tau$ is the eigen time.
 
The metric of space time has mass dimension $0$. Since the curvature is a second derivative, is has mass
dimension $2$. But the energy momentum tensor in three spacelike dimensions has dimension $1+3$. To get the
dimensions right in Eq.(\ref{EFG}), we thus have to introduce a mass scale $m_{\rm p}$ with a power of $2$, the so called Planck scale.

The exterior metric of an arbitrary spherical symmetric mass distribution with mass $M$ 
is described by the Schwarzschild solution \cite{Schwarzschild}:
\begin{eqnarray} \label{ssmetr}
{\mathrm{d}}s^2= g_{\mu \nu} {\rm d}x^{\mu}  {\rm d}x^{\nu} = -\gamma(r) {\mathrm{d}}t^2 + 
\gamma(r)^{-1}{\mathrm{d}}r^2 + r^2 {\mathrm{d}} \Omega^2 \quad ,
\end{eqnarray}
where 
\begin{eqnarray}  
\gamma(r)=1- \frac{1}{m_{\rm p}^2}\frac{2M}{r} \quad , \label{gamma}
\end{eqnarray}
and ${\mathrm{d}} \Omega$ is the surface element of the 3-dimensional unit sphere. 
 
To identify the constant $m_{\rm p}$, we take the limit to the familiar Newtonian potential by
considering a slow motion in a static gravitational field. Here, slow means
slow relative to the speed of light in which case the eigen time is almost identical to
the coordinate time, ${\rm d}t/{\rm d}\tau \approx 1$, and for the spatial
coordinates $i=1,2,3$ it is ${\rm d}x^i /{\rm d}\tau \approx 0$ in Eq.(\ref{Geo}). This yields
the approximation 
\begin{eqnarray}
\frac{{\mathrm{d}}^2x^{i}}{{\mathrm{d}}\tau^2}
&\approx&
\frac{{\mathrm{d}}^2x^{i}}{{\mathrm{d}}t^2}\approx -\Gamma^i_{tt}
\left(\frac{{\mathrm{d}}t}{{\mathrm{d}}\tau}\right)^2  
\approx-\Gamma^i_{tt} =g^{ij}\Gamma_{jtt}
\quad .
\end{eqnarray}
Using the definition of the Christoffel-symbols Eq.(\ref{Christoffel}) and taking into account the
time independence of the gravitational field, $\partial_t g_{\mu \nu}\approx 0$, we obtain 
\begin{eqnarray}
\frac{{\mathrm{d}}^2x^{i}}{{\mathrm{d}}t^2}\approx-\frac{1}{2}g_{tt,i} \; \; .
\end{eqnarray}
Inserting Eq. (\ref{gamma}) then yields
\begin{eqnarray}
\frac{{\mathrm{d}}^2x^{i}}{{\mathrm{d}}t^2}\approx  -\frac{M}{m_{\rm p}^2}\frac{1}{r^2} \quad,
\end{eqnarray} 
and we can identify the quantity $\gamma$ with the Newtonian Potential $\phi$ by
\begin{eqnarray}
\gamma(r)=1- 2 \phi(r) \quad .
\end{eqnarray}
So, it is $m_{\rm p}^2 = 1/G$, the inverse of the Newtonian constant, and the Planck mass turns out 
to be $\approx 10^{16}$~TeV. Compared to the typical
mass scales of the strong and electroweak interactions, which are $\approx 100$~GeV, this is a huge 
value and reflects the fact that gravity is much weaker. This large gap between the mass scales of the
Standard Model and gravity is also called the hierarchy problem.

Let us now come back to our test particle with a very high energy. 
Looking at Eq.(\ref{gamma}), we see that the energy $E$ inside
a space time volume with side length given by the Coulomb wave length $\lambda$, will cause a perturbation of 
the metric which is of order $E^2/m_{\rm p}^2$. Since flat space is described by $\gamma =1$, we can therefore draw 
the very important conclusion that a particle
with an energy close to the Planck mass will cause a non negligible perturbation of space time on length
scales comparable to its own Compton wavelength. Such a particle then should be subject to the yet
unknown theory
of quantum gravity. 
Physics at the Planck scale thus represents a future challenge, located between particle physics 
and General Relativity.

\section{Black Holes}
 
Black holes are an immediate consequence of the Schwarzschild solution Eq.(\ref{ssmetr}). 
As we see, the metric component $g_{tt}=\gamma(r)$ has a zero at the 
radius $R_H = 2 M/m_{\rm p}^2$, the so called Schwarzschild radius. 
For usual stellar objects like our sun is, this radius lies well inside the mass distributions where 
the exterior solution does no longer apply\footnote{The Schwarzschild radius for the sun is $\approx 3$~km, whereas
its actual radius is $\approx 100,000$~km.}. However, if the mass is so densely packed that it has a radius
smaller than $R_H$, we will be confronted with the consequences of $g_{tt}$ having a zero. 

The effect
is described most effective by looking at the redshift of a photon.
This photon is send out with frequency $\nu_0$ by an observer
at radius $r_{0}$ and is received at a radius $r_{1}>r_{0}$ with frequency $\nu_1$. 
To exclude effects due to Doppler shift, we assume both, sender and receiver, are in rest.
Frequencies count oscillations per time, so their
relation is inverse to the relation of the eigen time intervals at both locations   
\begin{eqnarray} \label{verh}
\frac{{\rm d} \tau_1}{{\rm d} \tau_0}=\frac{\nu_0}{\nu_1} \; \; .
\end{eqnarray}
Since both observers are in rest, it is ${\mathrm{d}}r^2={\mathrm{d}}\theta^2={\mathrm{d}}\phi^2=0$
on their world lines.
The relation between the eigen times, ${\rm d}\tau_{0/1}^2 = - {\rm d} s_{0/1}^2$, and the coordinate 
times is given by the metric in Eq.(\ref{ssmetr})
\begin{eqnarray}
{\mathrm{d}} \tau_0^2=\gamma(r_0){\mathrm{d}}t^2 \quad, 
\quad {\mathrm{d}}\tau_1^2=\gamma(r_1){\mathrm{d}}t^2
\quad .
\end{eqnarray}
By taking the square root and inserting the eigen time intervals into Eq.(\ref{verh}) we find that the 
photon will be received at the location $r_1$ with a frequency
\begin{eqnarray}
\nu_1 = \nu_0 \frac{\sqrt{\gamma(r_1)}}{\sqrt{\gamma(r_0)}} \quad.
\end{eqnarray}
 
This redshift goes to infinity for $r_0 \to R_H$, $\gamma(r_0) \to 0$. In this case, when the sender is close to
the Schwarzschild radius, the photon can not be sent out, no matter how high its initial energy
is. Thus, no information will ever reach the
observer at $r_1$. More generally speaking, it turns out that at $R_H$ the space time has a trapped
surface. What holds for the photon must also hold for slower moving objects.
Nothing can ever escape from the region inside the Schwarzschild radius. The object is completely black: a black hole. 
The radius $R_H$ is also called the event horizon.


\begin{parbox}[h]{14cm} 

\begin{figure}
\vspace*{-4.7cm}
\epsfig{figure=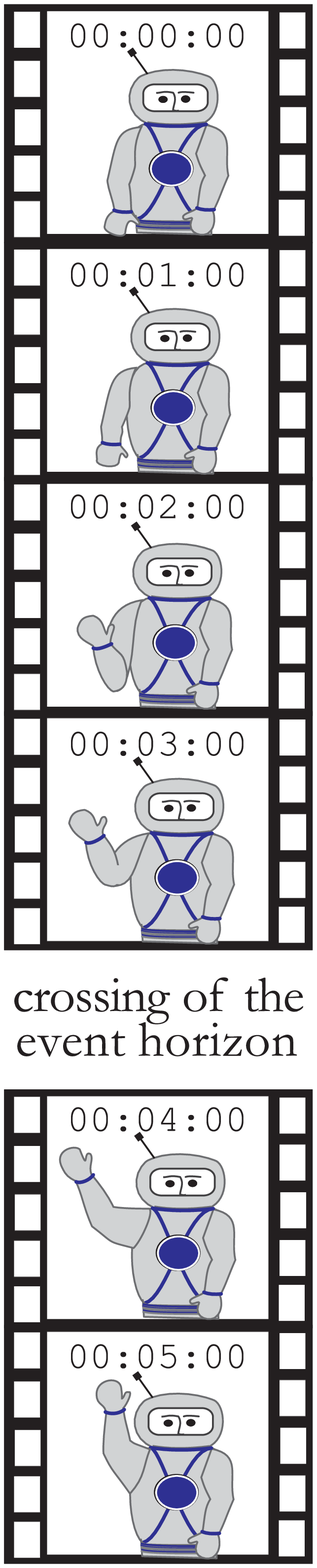, width=3cm }

\begin{center}\parbox{12cm}{
\caption{\label{astro} The last greetings from the astronaut who crosses the Schwarzschild radius. 
The left strip shows the pictures he sends out and the right strip shows the pictures that are 
received by an observer who is   far away from the horizon.}}\end{center}

\parbox{14cm}{
It is noteworthy at this point that the existence of a horizon, weird at it might seem, is not
troublesome in any regard. Even though the Schwarzschild metric in the above given form Eq.(\ref{ssmetr})
has a pole in the $g_{rr}$ component at $R_H$, this pole is not a physical one, meaning it can be
removed by a suitable coordinate transformation. All of the curvature components are perfectly well
behaved at the horizon. 
 } 

\vspace*{-20.0cm}
\hspace*{3.5cm}
\parbox{7cm}{The dependence of the eigen time intervals on the location
of the observer can be illustrated by examining the information that
is sent out by an infalling observer. Though the above derived expression
is not appropriate in this case (the observer is not located at a fixed
radius), the result is qualitatively the same, see e.g. \cite{Thermobh}.

The observer who approaches the horizon sends out information in time
intervals which are of equal length in his eigen time as depicted in
Figure \ref{astro}, left. These pictures are received by an observer
located at an arbitrary large radius. Here, space time is flat and
the coordinate time is identical to the eigen time. 

The pictures which are received by the observer far away from the horizon 
are shown in Figure \ref{astro}, right.
When looking at the last greetings from the infalling observer,
the receiver will notice that the time intervals the infalling observer claimed
to have equal length become longer and longer. Also, the wave lengths of the
photons become longer and the colors are red shifted as derived above. 
But in addition, it seems to the receiver that every motion of his
infalling friend becomes
slow motion until it eventually freezes when the sender
crosses the event horizon.

This of course does not only apply for infalling observers but also
for the collapsing matter. For this reason, black holes have also
been termed 'frozen stars'.
}

\vspace*{-15.2cm}
\hspace*{11cm}
\epsfig{figure=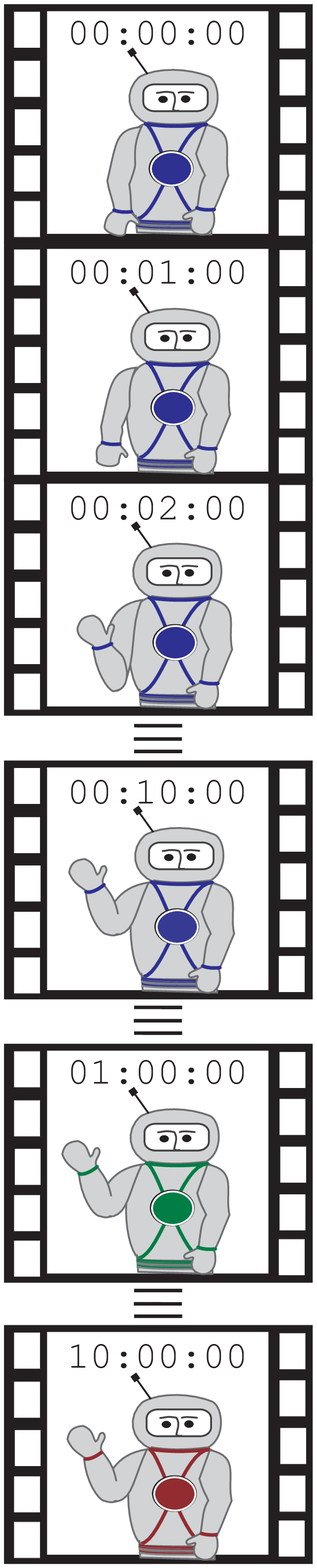, width=3cm }
\end{figure}
  
\end{parbox}
\pagebreak

What can not be removed is the singularity at $r=0$, which turns out to be
a point with infinite curvature. Infinities in physics usually signal that we have missed some
essential point in our mathematical treatment and, once again, indicate that the extreme regions of
gravity are poorly understood.

\subsection{Radiation of Black Holes}

In 1975, Hawking published his computation about the radiation of
black holes \cite{Hawk1}. He showed that a black hole can emit particles if
one takes into account the effects of quantum gravity in curved space time.
Since the not so black hole of mass $M$ is quasi-static (it is 'frozen'), the spectrum is a 
thermal spectrum and can be related to a temperature
\begin{eqnarray} \label{kappa}
T = \frac{\kappa}{2 \pi} \quad,\quad \mbox{with} \quad \kappa = \frac{1}{2} \partial_r \gamma(r)\big|_{r=R_H}\quad, 
\end{eqnarray}
where $\kappa$ is the surface gravity of the black hole.
If one inserts all constants one finds this temperature to be extremely small
\begin{eqnarray} \label{Hawktemp}
T 
= \frac{1}{8 \pi}\frac{c^2 m_{\rm p}^2}{k_B} \frac{1}{M} \approx 10^{-6}
\frac{M_{\odot}}{M} [\mathrm{K}] \quad. \label{temp}
\end{eqnarray}
Here, $M_{\odot}\approx 10^{54}$~TeV is the mass of the sun.

The smaller the mass of the black hole, the higher is its temperature.
It is remarkable that this formula for the first time connects all fundamental 
constants and joins gravitation, quantum field theory and thermodynamics.


\begin{parbox}[h]{15cm} 

\begin{figure}[h]
\framebox{\epsfig{figure=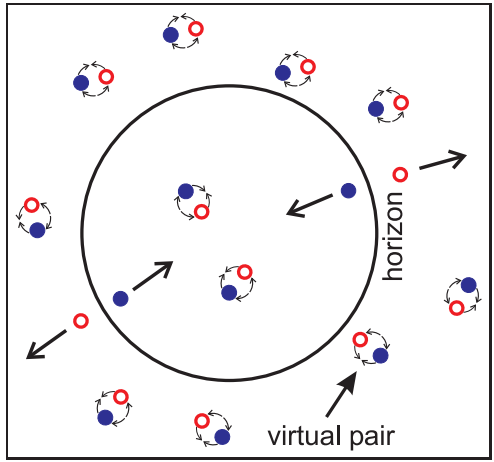, width=6.2cm }}

\parbox{6.5cm}{\begin{flushleft}
\caption{\label{parcre} Particle creation at the horizon.}
\end{flushleft}}

\vspace*{-7.9cm}
\hspace*{7cm}
\parbox{7cm}{
To understand this so called Hawking effect, we will use a simple analogy. 
The effect is similar to the particle production out of vacuum in quantum
electrodynamics. In both cases, a strong external field is present. The vacuum is thought
to be filled with virtual pairs of particles. In a background field strong enough, the
virtual pairs can be ripped apart and form a real pair. The energy for this pair
creation is provided by the background field. 
Figure \ref{parcre} shows a schematic picture of this process.
These analogies of course should not be put too far and can not 
replace a proper calculation. But they provide us with
a phenomenological understanding of the effect. 
} 
\end{figure}
 
\end{parbox}
 

One should be aware of two
crucial differences to the electromagnetic case. 
One essential difference is
that, in contrast to the electromagnetic case, in the gravitational case not the
actual strength of the field is responsible for the pairs becoming real but its tidal
forces are. This is due to the fact that the 'charges' of gravity are always positive.
The second weak point of this analogy is that the energy of the gravitational
field is not a well defined quantity. Since gravity is build on the fact that a local
observer can not distinguish between gravity and acceleration, this energy is a 
coordinate dependent quantity and can e.g. be defined in the asymptotically flat
coordinate system. In this system, the energy of the infalling particle is apparently
negative (this is due to $g_{tt}$ being negative, see Eq. (\ref{ssmetr})). The black hole 
therefore looses
mass when one of the particles escapes and the other one falls in, as we would have expected.

This picture will be sufficient for our purposes.
More details on the calculation can be found e.g. in \cite{Wald2,Novi,BiDa}.
  
The thermodynamic of black holes \cite{Thermobh,Wald2}  can be put even further and reveals amazing
parallels by relating geometrical quantities of the hole to thermodynamic variables. The most
important of this quantities is the entropy $S$. 

We know from the standard thermodynamics that the integrating factor   of the
entropy is the inverse of the temperature. By identifying the total energy of the
system with the mass of the black hole we find 
\begin{eqnarray}
\frac{\partial S}{\partial M} = \frac{1}{T} \quad.
\end{eqnarray}
After inserting Eq.  (\ref{kappa}) 
\begin{eqnarray}
\frac{\partial S}{\partial M} = 8 \pi \frac{M}{m_{\rm p}^2}  
\end{eqnarray}
and integrating we obtain
\begin{eqnarray}  
S(M)&=&4 \pi \left(\frac{M}{m_{\rm p}}\right)^2 + {\mbox{const.}} \nonumber \\
&=& \frac{m_{\rm p}^2 }{4} {\cal{A}} + 
{\mbox{const.}} \quad, \label{entropie4}
\end{eqnarray}
where ${\cal{A}} = 4 \pi R_H^2$ is the surface of the black hole. The additive constant is not
relevant for our further treatment and we will set it to zero.
 
The black hole looses mass through the evaporation process. Eq. (\ref{Hawktemp}) then states
that the temperature will be increased by this, resulting in an even faster mass loss. Using
the above results, we can estimate the lifetime of the black hole under the assumption
that it is a slow process. With all constants, the energy density 
of the radiation is then given by Stefan-Boltzmann's law 
\begin{eqnarray}
\varepsilon=\frac{\pi^2}{30 \hbar^3 c^2} (k_B T_S)^4 \;\;.
\end{eqnarray}
The variation of the mass of the black hole itself is then 
\begin{eqnarray} \label{Stefbo}
\frac{{\mathrm{d}}M}{{\mathrm{d}}t} \approx - \varepsilon {\cal A} = 
- \frac{1}{15\pi 8^3} \frac{c^4\hbar}{G^2} \frac{1}{M_0^2} \quad,
\end{eqnarray}
where $M_0$ is the initial mass of the black hole under investigation.
By integration over $t$ we find for the lifetime 
\begin{eqnarray} \label{lebenhawk}
\tau \approx {15\pi 8^3} \frac{G^2}{c^4\hbar} M_0^3 \approx 10^{59} 
\left(\frac{M_0}{M_{\odot}}\right)^3 \mathrm{Gyr} \quad,
\end{eqnarray}
after which the mass $M$ is reduced to 0. 

To get a grip on this quantity for astrophysical objects, let us insert some
numbers. For black holes caused by collapse of usual astrophysical objects,  
it is $M_0 > 10^{15}$~g and $\tau$ exceeds the age of the universe 
($\approx 10^{10}$ years). For super massive black holes, caused by secondary collapse 
of whole star clusters with typical masses of $M_0 \geq 10^{27}$~g, the temperature
is even below the 3K cosmic background radiation and it does not evaporate anyhow.
Mini black holes, which can be be created by density fluctuations in the early universe
would evaporate right now. The non-observation of this effects then sets limits on the
spectrum of the primordial density fluctuations.
 
Looking back to our calculation we will notice that
${\rm d}M/{\rm d}t$ goes to infinity for late times. This unphysical result
is due to the assumption that
the black hole is treated as a heat bath and the back reaction which modifies the
temperature by emission of a particle has been neglected. This estimation is pretty good for large and
cold objects, such as astrophysical black holes. In general, however, this 
will no longer be the case if the mass of the black hole itself is close to its 
temperature. The late stages of evaporation will then be modified. We should not
be surprised that the scale at which this will happen is when the mass of the
black hole is close to the Planck mass, as can been seen from Eq. (\ref{Hawktemp}).
 
By applying the laws of statistical mechanics we can improve the calculation in
such a way that it remains valid from the thermodynamical point of view even if the
mass of the emitted particle gets close to the mass of the black hole. This can be
done by using the micro canonical ensemble, as has been pointed out by Harms \cite{Harms}.

The Hawking result could also have been derived by using the canonical ensemble in which
the number density of the particles (here, bosons) is given by  
\begin{equation} \label{nkan}
n(\omega) = 
\frac{1}{\exp{\frac{\hbar \omega}{k_B T}}-1} \quad .
\end{equation}
As familiar from the case of black body radiation this yields the total energy which
we have used in (\ref{Stefbo}) by integration over the $3$ dimensional momentum space
\begin{eqnarray} \label{epsil}
\varepsilon = 
\frac{4 \pi}{(2 \pi)^3} 
\int_0^{\infty} n(\omega) \omega^3 {\mathrm d\omega} \propto T^4 \;\;.
\end{eqnarray} 
Here again, we use that the mass $M$ is much larger than the energy of the emitted particles.
Strictly spoken, it should not be possible to emit particles with $\omega>M$.

Now, instead of assuming the black hole being a heat bath, we use the 
micro canonical ensemble. Then, the number density for a single particle
with energy $\omega$ and $\omega \leq M$ is
\begin{equation} \label{nmik}
n(\omega) =  \frac{\exp{[S(M-\omega)]}}{{\rm exp}[S(M)]}\quad,
\end{equation}
where $S$ is the entropy derived in (\ref{entropie4}). 
 
The multi particle number density is
\begin{equation} \label{mehrteil}
n(\omega) = 
\sum_{j=1}^{\lfloor \frac{M}{\omega} \rfloor} \frac{{\rm exp}[S(M-j\omega)]}{{\rm exp}[S(M)]} \quad,
\end{equation}
where $\lfloor a \rfloor$ is the next smaller integer to $a$ and assures that nothing can be
emitted with energy above the mass of the black hole itself.
This yields then the total number density
\begin{eqnarray}
\varepsilon = \frac{4 \pi}{(2 \pi)^3} 
{\mathrm e}^{-S(M)} \int_0^{\infty}  \sum_{j=1}^{\lfloor \frac{M}{\omega} \rfloor}   
\exp{[S(M-j\omega)]} \omega^3 {\mathrm d}\omega \quad.
\end{eqnarray} 
  
We further substitute $x=(M-j\omega)$ (see also \cite{Page}). $x$ is then the energy after emission of
$j$ particles of energy $\omega$. After some algebra one finds
\begin{eqnarray}
\varepsilon = \frac{4 \pi}{(2 \pi)^3} 
{\mathrm e}^{-S(M)} \sum_{j=1}^{\infty} \frac{1}{j^4} \int_0^{M} {\mathrm e}^{S(x)} (M-x)^3 {\mathrm d}x\quad.
\end{eqnarray} 
The sum has the value $\pi^4 /90$. From this, the time dependence of the mass is given by
\begin{eqnarray}
\frac{{\mathrm d}M}{{\mathrm d}t} = \frac{4 \pi^3}{45}\frac{M^2}{m_{\rm p}^4} \; 
{\rm exp}{[-4\pi (M/m_{\rm p})^2]}  
\int_{0}^{M} (M-x)^{3} {\rm exp}{[4 \pi (x/m_{\rm p})^2]} {\mathrm d}x \quad.
\end{eqnarray}


\begin{parbox}{14cm} 

\vspace*{-1.0cm}
\begin{figure}[h]
\hspace*{-0.3cm}
\epsfig{figure=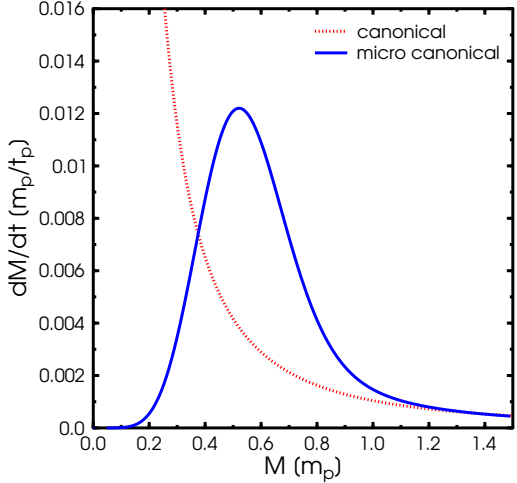, width=8.0cm}
 
\parbox{7cm}{\caption{\label{evap4} Canonical contra micro canonical.}}

\vspace*{-7.6cm}
\hspace*{7.3cm}
\parbox{6.7cm}{
Figure \ref{evap4}. shows this result in contrast to the canonical one for masses
close to $m_{\rm p}$. Here, $t_{\rm p}$ is
the timescale associated with the Planck mass $t_{\rm p}=\hbar /c^2 m_{\rm p}$. As can be seen, 
for large $M$, both cases agree. For small $M$, the canonical evaporation rate 
diverges whereas the micro canonical evaporation rate drops for $M<m_{\rm p}$ after passing
a maximum close by $m_{\rm p}$. 

By integrating over ${\mathrm d}M/{\mathrm d}t$, we can again compute the lifetime.
As a numerical example we investigate an
initial mass of $M=2m_{\rm p}$. Here, the micro canonical approach yields
$\tau \approx 1.45 \times 10^{17} \;{m_{\rm p}^{-1}} \approx \; 10^{-3} {\rm fm}/c$, whereas the
canonical approach would yield $\tau \approx 2 \times 10^6 \;{m_{\rm p}^{-1}} \approx \; 10^{-14}  {\rm fm}/c$.
} 

\end{figure} 
 
\end{parbox}


One might raise the objection that at the mass scales under consideration here, the effects of
quantum gravity should play the most important role. Nevertheless, thermodynamics for quantum systems
is a very powerful tool and unless we know more about quantum gravity we hope that it remains 
valid even close by the Planck scale. 
 
In this section we have only considered the case in which 
the black hole is completely determined by the mass trapped in it. It turns out that even 
the most general vacuum solution of Einsteins Field Equations depends on only three parameters. Besides the
mass $M$ of the black hole, it can also carry angular momentum $J$ and electric charge $Q$. The
metric for this system is known as the Kerr-Newman metric \cite{Kerr}, and the fact of the black hole solution only
having a three parameter set is known as the no-hair theorem.

The result that every black hole is described by this three parameters means that during
the collapse all other characteristics of the initial state, such as gravitational monopole moments, 
have to be radiated away. After this  balding phase, the hole remains with its last
three hairs: $M$, $J$ and $Q$.

The angular momentum and the charge do also modify the evaporation spectrum which can be included
in a straight forward way (see, e.g. \cite{Novi,Kanti:2004nr}). In case the initial state has angular momentum,
the emitted radiation will carry away most of its rotational energy in a spin down phase.

\section{Extra Dimensions}

So far it might seem that Planck scale physics, though interesting, is only important at such
high energies that it is out of reach for experimental examination on earth. This huge value of 
the Planck scale, however, might only be an apparent scale, caused by the presence of 
additional space like dimensions \cite{antoni}. In contrast to our usual three space like
dimensions, these additional dimensions are compactified to small radii which explains why we have
not yet noticed them.

Motivated by String Theory, the proposed models with extra dimensions lower the
Planck scale to values soon accessible.
These models predict a vast number of quantum gravity effects at the lowered
Planck scale, among them the production of TeV-mass 
black holes and gravitons.

\subsection{The Model of Large Extra Dimensions}
 
During the last decade, several models using compactified Large Extra Dimensions ({\sc LXD}s) as an
additional assumption to the quantum field theories of the Standard Model (SM) have
been proposed. The setup of these effective models is motivated by String Theory though
the question whether our spacetime has additional dimensions is well-founded on its own and
worth the effort of examination.

The models with {\sc LXD}s provide
us with an useful description to predict first effects beyond the SM.
They do by no means claim to be a theory of first principles or a candidate for a grand
unification. Instead, their simplified framework allows the derivation of 
testable results which can in turn help us to gain insights about the underlying theory.

There are different ways to build a model of extra dimensional space-time. Here, we want to
mention only the most common ones:
\begin{enumerate}
\item The {\sc ADD}-model proposed by Arkani-Hamed, Dimopoulos and Dvali \cite{add} adds $d$ extra
spacelike dimensions without curvature, in general each of them compactified to the same radius $R$. All 
SM particles are confined to our brane, while gravitons are allowed to propagate freely in the bulk. \label{ADD}
\item The setting of the model from Randall and Sundrum \cite{rs1,rs2} is a 5-dimensional spacetime with
an non-factorizable geometry. The solution for the metric is found by analyzing the solution of Einsteins 
field equations with an energy density on our brane, where the SM particles live. In the type I model \cite{rs1} 
the extra dimension is compactified, in the type II model \cite{rs2} it is infinite.
\item Within the model of universal extra dimensions \cite{uxds}
all particles (or in some extensions, only bosons) can propagate in the 
whole multi-dimen\-sional spacetime. The extra dimensions are compactified on an orbifold to 
reproduce SM gauge degrees of freedom.
\end{enumerate}
In the following we will focus on the model \ref{ADD}. which yields a beautiful and simple explanation
of the hierarchy problem. Consider a particle of mass $M$ located in a space time with $d+3$ dimensions.
The general solution of Poisson's equation yields its potential as a function of the radial distance $r$ to the
source
\begin{eqnarray}
\phi(r) \propto \frac{1}{M_{\rm f}^{d+2}} \frac{M}{r^{d+1}} \quad, \label{newtond}
\end{eqnarray}
where we have introduced a new fundamental mass-scale\footnote{Similar to our argument in
Section 1, we note that the curvature still has dimension $2$ but the energy momentum tensor now
has dimension $1+3+d$. To get the
dimensions right, we then have to introduce a mass scale with a power of $d+2$.}
$M_{\rm f}$.  
The hierarchy problem then is the
question why, for $d=0$, this mass-scale is the Planck mass, $m_{\rm p}$, and by a factor $10^{16}$ smaller
than the mass-scales in the SM, e.g. the weak scale.
    

\begin{figure}[h]
 
\epsfig{figure=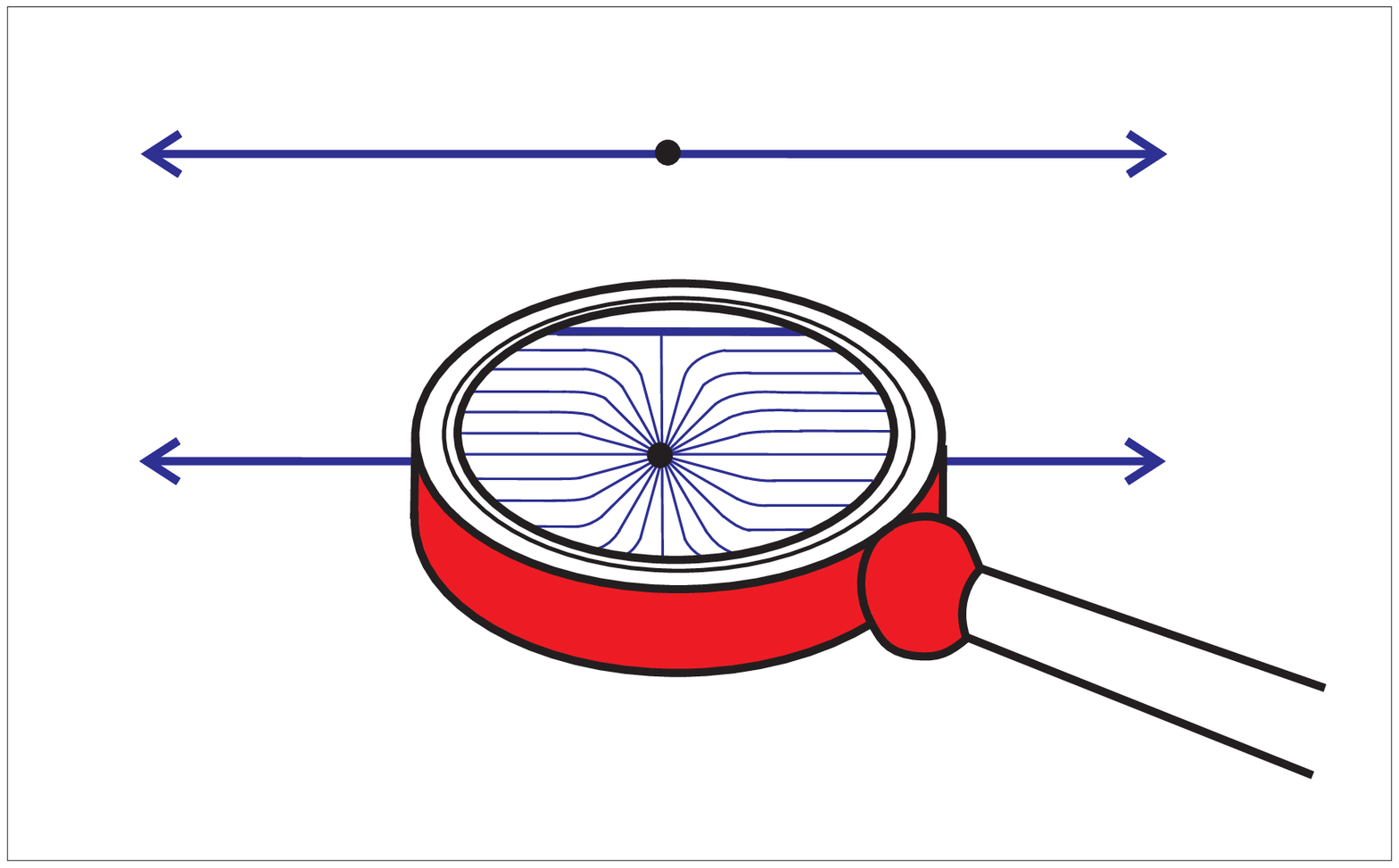,width=7.0cm}
\epsfig{figure=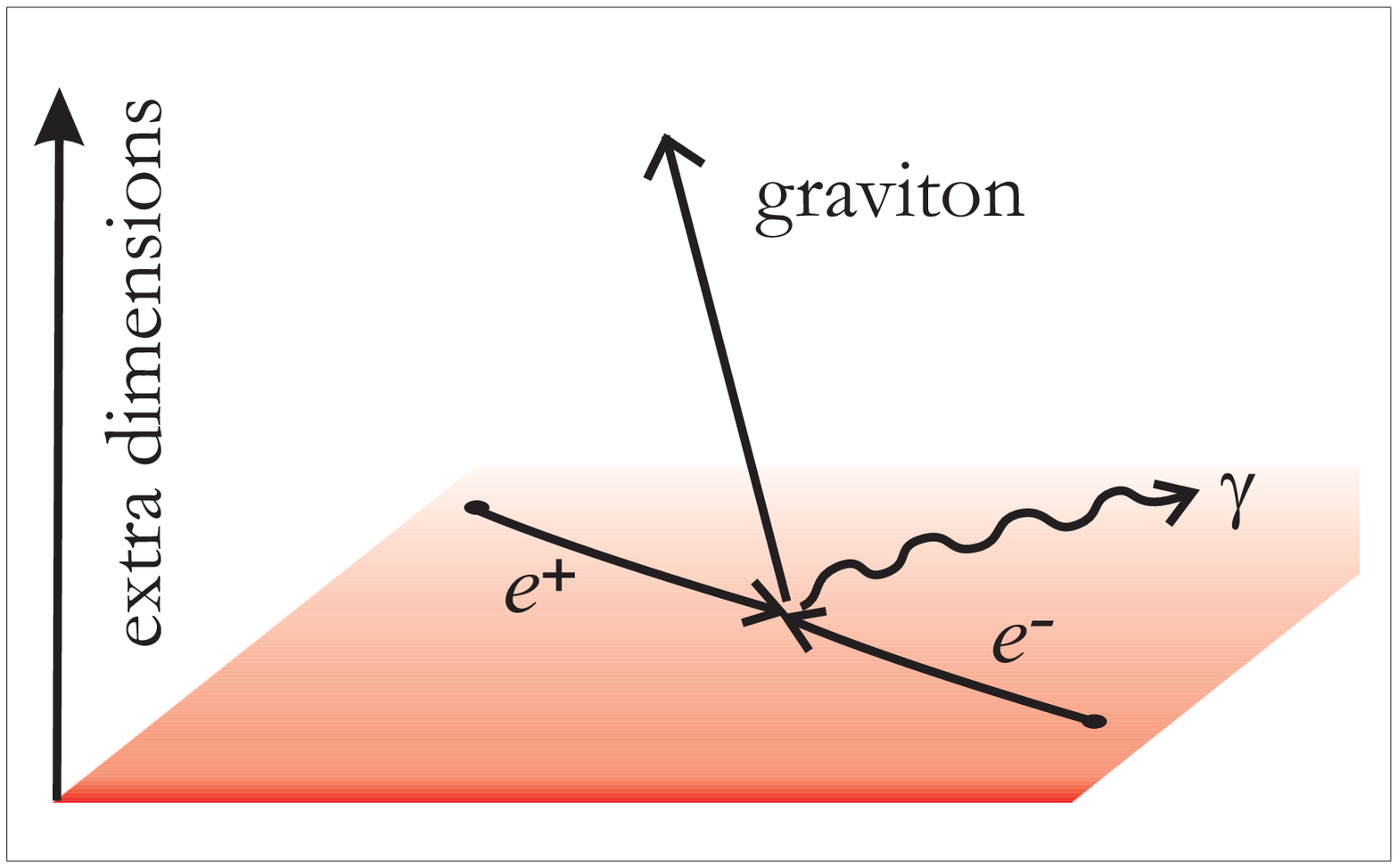,width=7.0cm}

\begin{center}\parbox{12cm}{
\caption{\label{grav}Left: at small distances the gravitational potential is higher dimensional. At
large distances, we rediscover the three dimensional case. Right: schematic figure for a scattering
process producing a graviton that escapes from our 3-dimensional submanifold and results in an energy loss.}}
\end{center} 
\end{figure}


The additional $d$ spacetime dimensions are compactified on radii $R$, which are
small enough to have been unobserved so far. Then, at distances $r \gg R$, the extra dimensions
will 'freeze out' and the potential Eq. (\ref{newtond})
will turn into the common $1/r$ potential, but with a fore-factor given by the volume of the extra 
dimensions
\begin{eqnarray}
\phi(r) \to \frac{1}{M_{\rm f}^{d+2}} \frac{1}{R^{d}} \frac{M}{r} \quad. \label{newton} 
\end{eqnarray}
This is also sketched in Figure \ref{grav}.
In the limit of large distances, we will rediscover the usual gravitational law which yields the relation
\begin{eqnarray}
m_{\rm p}^2 = M_{\rm f}^{d+2} R^d \quad. \label{Master}
\end{eqnarray}
Given that $M_{\rm f}$ has the right order of magnitude to be compatible with the other observed scales,
it can be seen from this argument that the volume of the extra dimensions suppresses the 
fundamental scale and thus, explains the huge value of the Planck mass.

The radius $R$ of these extra dimensions, for $M_{\rm f}\sim$~TeV, can be estimated with Eq.(\ref{Master}) and
typically lies in the range from $10^{-1}$~mm to $10^3$~fm for $d$ from $2$ to $7$, or the inverse radius 
$1/R$ lies in energy range eV to MeV, respectively. The case $d=1$ is excluded. It would result in
an extra dimension about the size of the solar system. 

Due to the compactification, momenta in the direction of the {\sc LXD}s can only occur in quantized steps 
{$\propto 1/R$} for every particle which is allowed to enter the bulk. The fields can be expanded in
Fourier-series
\begin{eqnarray}
\psi(x,y) = \sum_{n=-\infty}^{+\infty} \psi^{(n)}(x) \exp\left( {\rm i} n y/R \right) \quad,
\end{eqnarray} 
where $x$ are the coordinates on our brane and $y$ the coordinates of the {\sc LXD}s. 
This yields an infinite number of equally spaced excitations, the so called Kaluza-Klein-Tower.
On our brane, these massless KK-excitations act like massive particles, since the
momentum in the extra dimensions generates an apparent mass term
\begin{eqnarray}
\left[ \partial_x\partial^x - \left(\frac{n}{R}\right)^2 \right] \psi^{(n)}(x) = 0\quad.
\end{eqnarray} 

\subsection{Observables of Extra Dimensions}

The most obvious experimental test for the existence of extra dimensions is a 
measurement of the Newtonian potential at sub-mm distances. Cavendish like experiments which search
for deviations from the $1/r$ potential have been performed during the last years with high
precision \cite{Newtonslaw} and require the extra dimensions to have radii not larger than $\sim 100\mu$m,
which disfavors the case of $d=2$.

Also the consequences for high energy experiments are intriguing.
Since the masses of the KK-modes are so low, they get excited easily but it is not until
energies of order $M_{\rm f}$ that their phase-space makes them give an important contribution
in scattering processes.
The number of excitations $N(\sqrt{s})$ below an energy $\sqrt{s}$ 
can, for an almost
continuous spectrum, be estimated by integration over the volume of the $d$-
dimensional sphere whose radius is the maximal possible wave number $n_{\rm max}= R\sqrt{s}$
\begin{eqnarray}
N(\sqrt{s}) = \int_0^{n_{\rm max}} {\rm d}^d n = \Omega_{(d)} (R \sqrt{s})^d \quad. 
\end{eqnarray} 
 We can then 
estimate the total cross-section for a point interaction, e.g. $e^+ e^- \to {\rm graviton} + \gamma$ 
(depicted in Figure \ref{bumm}, right) by
\begin{eqnarray}
\sigma(e^+ e^- \to G \gamma) \approx \frac{\alpha}{m_{\rm p}^2}  N(\sqrt{s}) \approx  
\frac{\alpha}{s} \left( \frac{\sqrt{s}}{M_{\rm f}}\right)^{d+2} \quad,
\end{eqnarray} 
where we have used Eq.(\ref{Master}) and $\alpha$ is the fine structure constant. As can be seen, at energy scales close to the new
fundamental scale, the estimated cross-section becomes comparable to cross-sections of
electroweak processes.

The necessary Feynman rules for exact calculations of
the graviton tree-level interactions have be derived \cite{Hewett:2002hv} and the cross-sections have
been examined closely. 
Since the gravitons are not detected, their emission would lead to an energy loss in the 
collision and to a higher number of monojets. Modifications of SM predictions do
also arise by virtual graviton exchange, which gives additional contributions in the calculation
of cross-sections.

Another exciting signature of {\sc LXD}s is the possibility of black hole production as will be
discussed in detail the next section. It is a direct consequence of the gravitational force being
modified at distances smaller than the radius of the extra dimension.

These fascinating processes could be tested at the Large Hadron Collider ({\sc LHC}), which is
currently under construction at {\sc CERN} and is scheduled to launch in September 2007. The
{\sc LHC} will collide proton beams with a c.o.m.\footnote{center of mass} energy of 
$\sqrt{s}\approx$14~TeV. 

The here described effects would not only be observable in high energy collisions in the
laboratory but also in cosmic ray events.
Ultra high energetic comic rays ({\sc UHECR}s) have been measured with energies up 
to $\approx 10^8$~TeV! Though this energy seems enormous, one has to keep in mind 
that the infalling {\sc UHECR} hits an atmospheric nuclei in rest, which corresponds to a 
fixed target experiment. To compare to 
the energies reachable at colliders, the energy has to be converted in the c.o.m. system. 
A particle with a typical energy of $E \sim 10^7$~TeV in the laboratory frame which interacts with a nuclei
of mass $m \sim 1$~GeV in rest corresponds to a c.o.m. energy $\approx \sqrt{2 E m } \approx 100$~TeV. 
Nevertheless, this energy is still high enough to make {\sc UHECR}s an important window to study 
first effects beyond the Standard Model. Possible observables have been examined e.g. in Refs. 
\cite{cosmicrayskk,cosmicraysbh}.
 
Further, the emission of 
gravitons and black hole production leads to astrophysical consequences, such as enhanced
cooling of supernovae and modification of the cosmic background radiation. The presently 
available data from collider physics as well as from astrophysics sets 
constraints on the parameters of the model. For a recent update see 
e.g. \cite{Cheung:2004ab}.

As we have seen, the {\sc LXD}-model predicts a rich phenomenology and is an useful extension of
the Standard Model that can be used to verify or falsify the existence of extra dimensions.

\section{Black Holes in Extra Dimensions}

In the standard
$3+1$ dimensional space-time, the production of black holes requires a concentration of 
energy-density which can not be reached in the laboratory. As we have seen, 
in the higher dimensional space-time, gravity becomes stronger at small distances and
therefore the event horizon is located at a larger radius. 

\begin{figure}
\centering
\setlength{\fboxsep}{3mm}
\framebox{\epsfig{figure=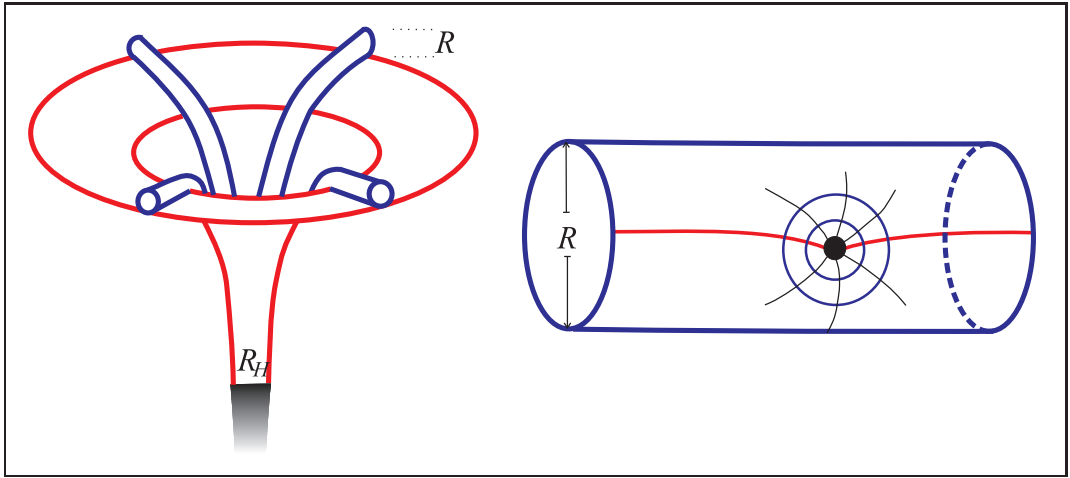  ,width=13.4cm}  
}
\begin{center}\parbox{12cm}{
\caption{Sketch of the black hole topology in a space time with compactified 
extra dimensions. The left picture shows the astrophysical black hole with $R_H \gg R$, the right
picture shows the collider produced black hole with $R_H \ll R$.}
\label{add}}
\end{center}
\end{figure}

However, for astrophysical objects we expect to find back the usual 3-dimensional Schwarz\-schild
solution. In this case, the horizon radius is much larger than the radius of the extra dimensions
and the influence of the extra dimensions is negligible. 
We, in the contrary, will be interested in the case where the black hole has a mass close to
the new fundamental scale. This corresponds to a radius 
close to the inverse new fundamental scale, and thus $R_H \ll R$.   
Those two case are depicted in Figure \ref{add}.

In the case $R_H \ll R$ under investigation, the topology of the object can be assumed to
be spherical symmetric $d+3$ dimensions. The boundary conditions from the compactification
can be neglected. The hole is so small that it effectively does not notice the periodicity.


The higher dimensional Schwarzschild-metric has been derived in \cite{my} and takes the form
\begin{equation}
{\rm d}s^2 
=
- \gamma (r) {\mathrm d}t^2
+ \gamma^{-1}(r) {\mathrm d}r^2
+ r^2 {\mathrm d}\Omega^2_{(d+3)}
\quad ,
\end{equation}
where ${\rm d}\Omega_{(d+3)}$ now is the surface element of the 
$3+d$ - dimensional unit sphere and
\begin{equation}
\gamma(r)=1-\left(\frac{R_H}{r}\right)^{d+1} \quad.
\end{equation}
The constant $R_H$ again can be found by requiring that we reproduce the
Newtonian limit for $r\gg R$, that is $1/2 \partial_r \gamma$ has to 
yield the Newtonian potential:
\begin{eqnarray}
\frac{d+1}{2} \left(\frac{R_H}{r}\right)^{1+d} \frac{1}{r} = \frac{1}{M_{\rm f}^{d+2}} \frac{M}{r^{d+2}} \quad.
\end{eqnarray}
So we have
\begin{eqnarray}
\gamma(r) =  1- \frac{2}{d+1} \frac{1}{M_{\rm f}^{d+2}} \frac{M}{r^{d+1}} \quad,
\end{eqnarray}
and $R_H$ is\footnote{Note, that this does not agree with the result found in \cite{my} because in this
case the extra dimensions were not compactified and therefore the constants were not matched to
the four-dimensional ones. The fore-factors cancel in our case with the fore-factors from the
higher dimensional Newtonian law. Though this is not in agreement with the relation used in
most of the literature, it is actually the consequence of relation Eq. (\ref{Master}). These fore-factors, 
however, are of order one and can be absorbed in an redefinition of the new fundamental scale $M_{\rm f}$.}
\begin{equation} \label{ssradD}
R_H^{d+1}=
\frac{2}{d+1}\left(\frac{1}{M_{\rm f}}\right)^{d+1} \; \frac{M}{M_{\rm f}} \;\; .
\end{equation}
It is not surprising to see that a black hole
with a mass about the new fundamental mass $M \sim M_{\rm f}$, has a radius of about the new fundamental
length scale $L_{\rm f}=1/M_{\rm f}$ (which justifies the use of the limit $R_H\ll R$). 
For $M_{\rm f}\sim$~1TeV this radius is $\sim 10^{-4}$~fm. Thus, at the {\sc LHC} it
would be possible to bring particles closer together than their horizon. A black hole could be created.

The surface gravity  can be computed using Eq.(\ref{kappa})
\begin{equation} \label{kappaD}
\kappa= \frac{1+d}{2} \frac{1}{R_H} \quad,
\end{equation}
and the surface of the horizon is now given by 
\begin{eqnarray}
{\cal{A}} &=& \Omega_{(d+3)} R_H^{d+2} 
\\
&=& \Omega_{(d+3)} \frac{M}{\kappa}  \left( \frac{1}{M_{\rm f}}\right)^{d+2} 
\quad, \label{oberfl}
\end{eqnarray}
where $\Omega_{(d+3)}$ is the surface of the $d+3$-dimensional unit sphere.
\begin{eqnarray}
\Omega_{(d+3)} = \frac{2 \pi^{\frac{d+3}{2}}}{\Gamma({\frac{d+3}{2}})}\;\;.
\end{eqnarray}
 
  
\begin{parbox}[h]{14cm}    

\begin{figure}
  
\epsfig{figure=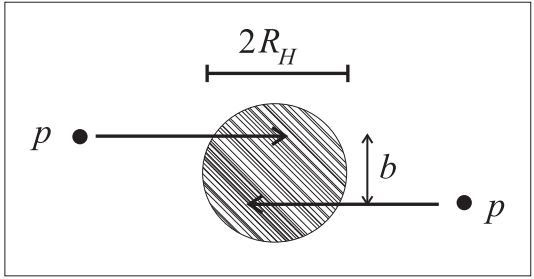,width=7cm }  

\vspace*{-4cm}
\hspace*{7.5cm}
\parbox{6.5cm}{
\caption{Schematic figure for a particle collision. At very high energies, the particles, $p$,
can come closer than the Schwarzschild radius, $R_H$, associated with their energy. If the
impact parameter, $b$, is sufficiently small, such a collision will inevitably generate
a trapped surface. A black hole has been created.}
\label{bumm}}
\end{figure}
\end{parbox}

\subsection{Production of Black Holes}

Let us consider two elementary particles, approaching each other with a very high kinetic energy in
the c.o.m. system close to the new 
fundamental scale $M_{\rm f}\sim 1$~TeV, 
as depicted in Figure \ref{bumm}. At those high energies, the particles can come
very close to each other since their high energy allows a tightly packed wave package
despite the uncertainty relation.
If the impact parameter is small enough, which will happen to a certain fraction of
the particles, we have the two particles plus their large kinetic energy in a very small
region of space time. If the region is smaller than the Schwarzschild radius connected
with the energy of the partons, the system will collapse and form a black hole.

The production of a black hole in a high energy collision is probably the most inelastic 
process one might think of. Since the black hole is not an ordinary particle of the Standard Model
and its correct quantum theoretical treatment is unknown, it is treated as a metastable
state, which is produced and decays according to the semi classical formalism of black
hole physics.

To compute the production details, the cross-section of the black holes can be approximated
by the classical geometric cross-section 
\begin{eqnarray} \label{cross}
\sigma(M)\approx \pi R_H^2 \quad,
\end{eqnarray}
an expression which does only contain the fundamental Planck scale as coupling constant. This  
cross section has been under debate \cite{Voloshin:2001fe}, 
but further investigations
justify the use of the classical limit at least up to energies of 
$\approx 10 M_{\rm f}$ \cite{Solodukhin:2002ui}. However, the topic is still under
discussion, see also the very recent contributions \cite{Rychkov:2004sf}.

A common approach to improve the naive picture of colliding point particles,
is to treat the creation of the horizon as a 
collision of two shock fronts in an Aichelburg-Sexl geometry describing the fast moving particles
 \cite{GrWav}. Due to the high velocity of the moving particles, space time
before and after the shocks is almost flat and the geometry 
can be examined for the occurrence of trapped surfaces. 

These semi classical considerations do also give rise to
form factors which take into account that not the whole initial energy is
captured behind the horizon. These factors have been calculated in \cite{Formfactors}, depend on the number of extra dimensions, and are of order one. For our further qualitative discussion we will neglect them. 
A time evolution of the system forming the black hole is not provided is this treatment\footnote{If 
we think about it, this is in general also not provided in Standard Model processes.} but
it has been shown that the naively expected classical result remains valid also in String Theory 
\cite{Polchi}. 
 
Looking at Figure \ref{bumm}, we also see that, due to conservation laws, the angular momentum 
$J$ of the formed object only vanishes in completely central collisions with zero
impact parameter, $b=0$. In the general case, we will have an angular momentum  
$J  \approx 1/2 M b$.  
However, this modification turns out to be  
only about a factor $2$ \cite{Solo}. In general, the black hole will also carry a charge which
gives rise to the exciting possibility of naked singularities. We will not further
consider this case here, the interested reader is referred to \cite{Casadio:2001wh}. 

Another assumption which goes into the production details is the existence of a 
threshold for the black hole formation. From General Relativistic arguments, 
two point like particles in a head on collision
with zero impact parameter will {\sl always} form a black hole, no matter how large or small their energy. At small
energies, however, we expect this to be impossible due to the smearing of the wave functions by
the uncertainty relation. This then results in a necessary minimal energy to allow for the required
close approach. This threshold is of order $M_{\rm f}$, though the exact value is unknown since quantum
gravity effects should play an important role for the wave functions of the colliding particles.
For simplicity, we will in the following set this threshold equal to $M_{\rm f}$ but we want to point out
that most of the analysis in the literature assumes a parameter of order one. 
 
This threshold is also important as it defines the window in which other effects of the
LXD scenario can be observed. Besides the graviton production and the excitation of 
KK-modes, also string excitations \cite{Burikham:2004su} have been examined. At 
even higher energies, highly exited long strings, the stringballs \cite{Dimopoulos:2001qe}, are
expected to dominate the scattering process. At the threshold energy, the so 
called correspondence point, the excited states can be seen either as strings 
or as black holes \cite{Horowitz:1996nw}. The smooth transition from strings to black 
holes has  recently been investigated in Ref. \cite{Veneziano:2004er}. The 
occurrence of these string excitations   
can increase the transverse size of the colliding objects and thus, make black 
hole production more 
difficult. This important point will be further discussed in Section \ref{MinimalLength}. 

Setting $M_{\rm f}\sim 1$TeV and $d=2$ one finds 
$\sigma \sim 400$~pb.
Using the geometrical cross section formula, it is now possible to compute the differential cross section
${\mathrm d}\sigma/{\mathrm d}M$ which will tell us how many black holes will be formed with a certain
mass $M$ at a c.o.m. energy $\sqrt{s}$. 
The probability that two colliding particles will form a black 
hole of mass
$M$ in a proton-proton collision at the {\sc LHC} involves the parton distribution functions. These
functions, $f_A(x,\hat{s})$, parametrize the probability of finding a constituent $A$ of the
proton (quark or gluon) with a momentum fraction $x$ of the total energy of the proton. 
These constituents are also called partons. Here,
$\sqrt{\hat{s}}$ is the c.o.m. energy of the parton-parton collision. 

\begin{figure}[h]
\centering
\epsfig{figure=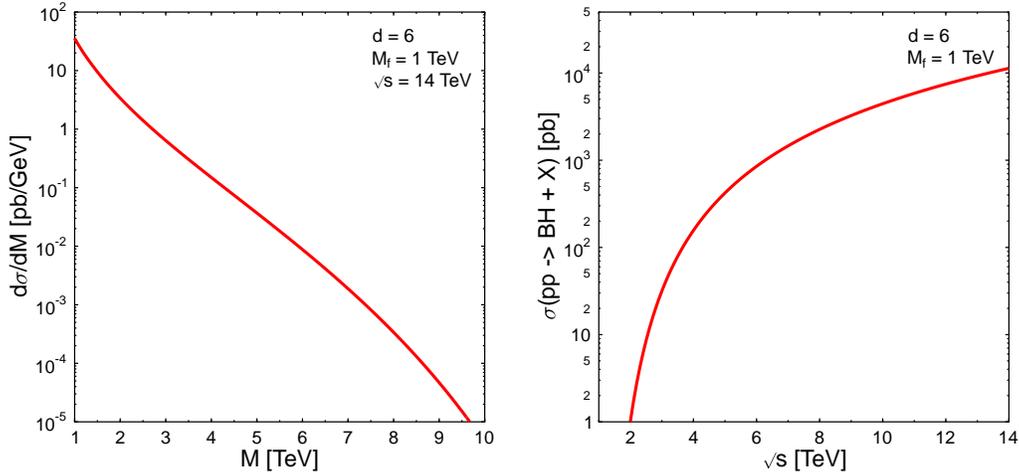, width=14cm}
 
\parbox{12cm}
{\caption{The left plot shows the differential cross section for black hole production in  
proton-proton-collisions at the {\sc LHC} for $M_{\rm f}=1$~TeV. The right plot shows the integrated total
cross section as a function of the collision energy $\sqrt{s}$. In both cases, the curves for various
$d$ differ from the above depicted ones by less than a factor 10.\label{dsdm}} }
 
\end{figure}

The differential cross section is then given by summation over all possible parton interactions and
integration over the momentum fractions, where the kinematic relation $x_1 x_2 s=\hat{s}=M^2$ has
to be fulfilled. This yields
\begin{eqnarray} \label{partcross}
\frac{{\rm d}\sigma}{{\rm d}M}  
&=&  \sum_{A_1, B_2} \int_{0}^{1} {\rm d} x_1 \frac{2 \sqrt{\hat{s}}}{x_1s} f_A(x_1,\hat{s}) 
f_B (x_2,\hat{s})  \sigma(M,d)   \quad.
\end{eqnarray}
The particle distribution functions for $f_A$ and $f_B$ are tabulated e.g. in the {\sc CTEQ} - 
tables. A numerical evaluation of this expression results in the differential
cross section displayed in Figure \ref{dsdm}, left. Most of the black holes
have masses close to the production threshold. This is due to the fact that at high collision
energies, or small distances respectively, the proton contents a high number of gluons which dominate the
scattering process and distribute the total energy among themselves.


It is now straightforward to compute the total cross section by integration over Eq. (\ref{partcross}), see 
Figure \ref{dsdm}, right. This also allows us to estimate the total number of black holes, $N_{\rm BH}$, 
that would be created at the {\sc LHC} per year. It is $N_{\rm BH}/{\rm year}= \sigma(pp \to {\rm BH}) L$
where we insert the estimated luminosity for the {\sc LHC},  $L=10^{33}{\rm cm}^{-2}{\rm s}^{-1}$. This
yields at a c.o.m. of $\sqrt{s}=14$~TeV a number of $N_{\rm BH} \approx 10^9$ per year! This means, about 
one black hole per second would be created. This result does also agree with the values found in \cite{dim}.
The importance of this process led to a high number of publications on the topic of TeV-mass black holes at
colliders \cite{dim,ehm,Giddings3,Hofmann:2001pz,Hossenfelder:2001dn,Cavaglia:2004jw,Atlas,BHex,BHlep}. 

The here presented calculation applies for the Large Hadron Collider and uses the parton distribution functions.
At such high energies, these functions are extrapolations from low energy measurements and one might
question whether the extrapolation can be applied into the regime of quantum gravity, 
or whether additional effects might modify them. Modifications might  e.g. arise through
the occurrence of a minimal length scale \cite{Hossenfelder:2004ze} or the presence of virtual graviton or 
KK-excitations. One should also keep in mind that the used extrapolations, the 
{\sc DGLAP}\footnote{Dokshitzer-Gribov-Lipatov-Altarelli-Parisi} equations \cite{DGLAP}, rely on
sliding scale arguments similar to the running of the coupling constants which is known to be modified
in the presence of extra dimensions \cite{Dienes:1998vg}. 

In this context it is therefore useful to examine black hole production at lepton colliders, 
which excludes the parton distribution functions as a source of uncertainty. A recent analysis has been given
by in \cite{BHlep} for a muon collider. In general, the  number for the expected black hole
production at lepton colliders will be higher than for hadron colliders at equal energies, which 
is due to the collision  energy not being distributed among the hadron constituents.

High energetic collision with energies close by or even above the new fundamental scale
could also be reached in cosmic ray events \cite{cosmicraysbh}. Horizontally
infalling neutrinos, which are scattered at nuclei in the atmosphere are expected to
give the most important and clearest signature. 
  
\subsection{Evaporation of Black Holes}
 
One of the primary observables in high energetic particle collisions is the transverse momentum of
the outgoing particles, $p_T$, the component of the momentum transverse to the direction of the beam.
Two colliding partons with high energy can produce a pair of outgoing particles, moving in opposite
directions with  high $p_T$ but carrying a color charge, as depicted in Figure \ref{bhphase}, top. 
Due to the quark confinement, the color has
to be neutralized. This results in a shower of several bound states, the hadrons, which includes
mesons (consisting of a quark and an antiquark, like the $\pi$'s) as well as baryons (consisting of three quarks, like
the neutron or the proton). The number of these
produced hadrons and their energy depends on the energy of the initial partons. This process
will cause a detector signal with a large number of hadrons inside a small opening angle. Such
an event is called a 'jet'. 
 
Typically these jets come in pairs of opposite direction. A smaller number of them can also be
observed with three or more outgoing showers. This observable will be strongly influenced by
the production of black holes.

To understand the signatures that are caused by the black holes we have to examine their evaporation
properties. As we have seen before, the smaller the black hole, the larger is its temperature and so,
the radiation of the discussed tiny black holes is the dominant signature caused by their presence.
As we see from Eq. (\ref{kappaD}) and Eq. (\ref{Hawktemp}) the typical temperature lies in the range of several $100$~GeV. Since most
of the particles in the black body radiation are emitted with this average energy, we can estimate the
total number of emitted particles to be of order $10$-$100$. As we will see, 
this high temperature results in a very short lifetime such that the black hole will
decay close by the collision region and can be interpreted as a metastable intermediate
state.    

\vspace*{1cm}

  
\begin{parbox}[h]{14cm}    

\begin{figure}
 
\setlength{\fboxsep}{2mm}
\framebox{
\epsfig{figure=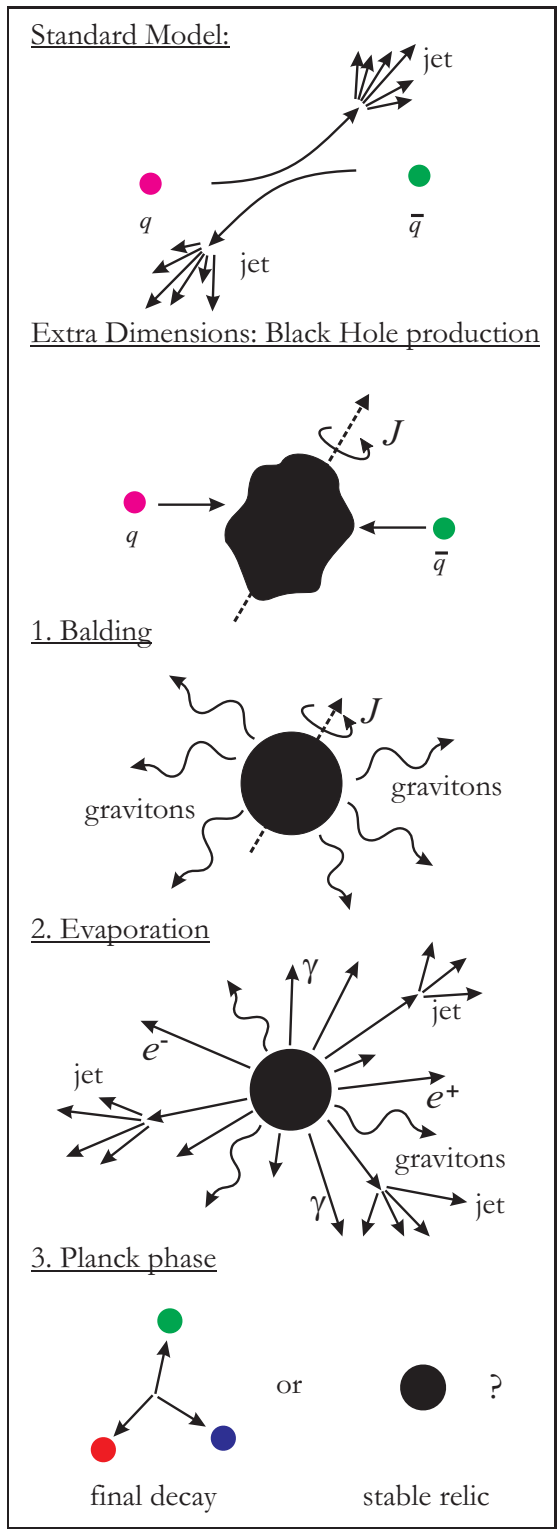, height=7.8in }}

\parbox{7cm}{\caption{\label{bhphase} Phases of black hole evaporation.}}

\vspace*{-8.3in}
\hspace*{7.5cm}
\parbox{6.5cm}{
Once produced, the black holes will undergo an evaporation process whose thermal
properties carry information about the parameters $M_{\rm f}$ and $d$. An analysis of the evaporation
will therefore offer the possibility to extract knowledge about the topology of our
space time and the underlying theory.

The evaporation process can be categorized in three characteristic stages \cite{Giddings3}, see
also the illustration in Figure \ref{bhphase}:
\bigskip 

1. {\sc Balding phase:} In this phase the black hole radiates away the multipole moments it has inherited
from the initial configuration, and settles down in
a hairless state. During this stage, a certain fraction of the initial mass will be lost in gravitational
radiation.
\bigskip 
 
2. {\sc Evaporation phase:} The evaporation phase starts with a spin down phase in which the Hawking
radiation carries away the angular momentum, after which it proceeds with emission of thermally
distributed quanta until the black hole reaches Planck mass. The radiation spectrum contains
all Standard Model particles, which are emitted on our brane, as well as gravitons, which are also
emitted into the extra dimensions. It is expected that most of the initial
energy is emitted in during this phase in Standard Model particles.  
\bigskip 

3. {\sc Planck phase:} Once the black hole has reached a mass close to the Planck mass, it falls into
the regime of quantum gravity and predictions become increasingly difficult. It is generally
assumed that the black hole will either completely decay in some last few Standard Model particles or
a stable remnant will be left, which carries away the remaining energy. 
 
}
 
\end{figure}
\end{parbox}

\begin{figure}[h]
\centering
\epsfig{figure=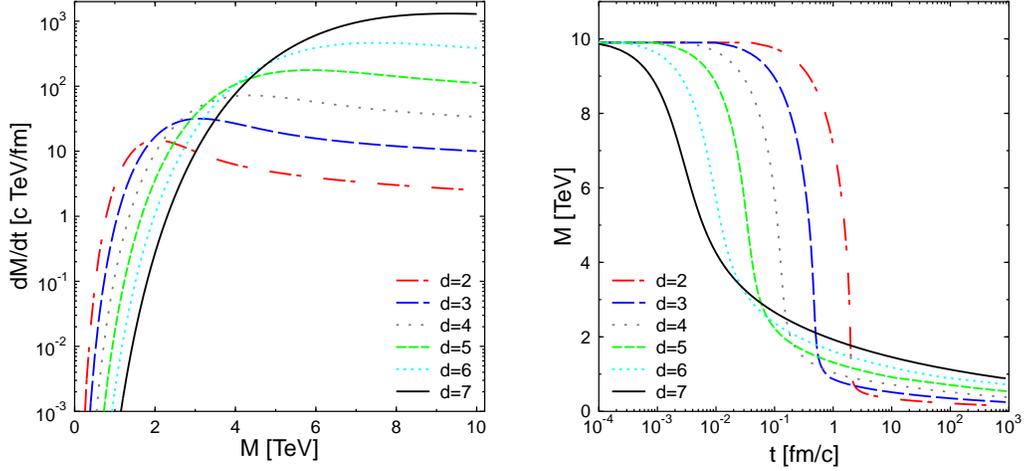,width=14cm}

\parbox{12cm}{\caption{The left plot shows the evaporation rate in as 
a function of the initial mass for various $d$. The right plot shows the
time evolution of a black hole with an initial mass of 10~TeV. \label{mdot}}}
 
\end{figure}

The evaporation rate ${\rm d}M/{\rm d}t$ of the higher dimensional black hole can be
computed using the thermodynamics of black holes.
With Eq. (\ref{kappaD}), (\ref{oberfl}) and (\ref{kappa}) we find for the temperature of
 the black hole  
\begin{eqnarray} \label{tempD}
T=\frac{1+d}{4 \pi}\frac{1}{R_H} \quad,
\end{eqnarray}
where $R_H$ is a function of $M$ by Eq. (\ref{ssradD}). Integration of the inverse temperature over $M$ then
yields  (compare to Eq.(\ref{entropie4})) 
\begin{eqnarray}  
S(M)
&=& 2 \pi  \frac{d+1}{d+2} \left( M_{\rm f} R_H \right)^{d+2}\quad.
\end{eqnarray}
For the spectral energy density we use the micro canonical particle 
spectra Eq. (\ref{nmik}) and (\ref{mehrteil}) but we now have to integrate over 
the higher dimensional momentum space
\begin{eqnarray}
\varepsilon = \frac{\Omega_{(d+3)}}{(2 \pi)^{3+d}} 
{\mathrm e}^{-S(M)} \sum_{j=1}^{\infty} \frac{1}{j^{d+4}} \int_0^{M} {\mathrm e}^{S(x)} (M-x)^{3+d} {\mathrm d}x
\quad,
\end{eqnarray}
where the value of the sum is given by a $\zeta$-function. From this we obtain the evaporation rate
\begin{eqnarray} \label{mdoteq}
\frac{{\mathrm d}M}{{\mathrm d}t} = \frac{\Omega_{(d+3)}^2}{(2\pi)^{d+3}} R_H^{2+d} \zeta(4+d) \; 
{\rm e}^{-S(M)}  
\int_{0}^{M} (M-x)^{(3+d)} {\rm e}^{S(x)} {\mathrm d}x \quad.
\end{eqnarray}
A plot of this quantity for various $d$ is shown in Figure \ref{mdot}, left. The mass dependence resulting from it
is shown in Figure \ref{mdot}, right. We see that the evaporation process slows down in the late stages and
enhances the life time of the black hole. 


To perform a realistic simulation of the evaporation process, one has to take into account 
the various particles of the Standard Model with the corresponding degrees of freedom and
spin statistics. In the extra dimensional scenario, Standard Model particles are bound to
our submanifold whereas the gravitons are allowed to enter all dimensions. It has been argued
that black holes emit mainly on the brane \cite{ehm}. The ratio between the evaporation in
the bulk and in the brane can be estimated by applying Eq. (\ref{mdoteq}) to $3$ or $3+d$ dimensions,
respectively. For simplicity we will assume that the mass is high enough such that the
result approximately agrees with the higher dimensional Stefan-Bolzmann law. Then it is
\begin{eqnarray}  
\frac{{\mathrm d}M}{{\mathrm d}t} &=& \frac{\Omega_{(3)}^2}{(2\pi)^{3}} R_H^{2} \zeta(4) \Gamma(4) \; 
T^4 \quad {\mbox{on the brane, and}}\\
\frac{{\mathrm d}M}{{\mathrm d}t} &=& \frac{\Omega_{(3+d)}^2}{(2\pi)^{3+d}} R_H^{2+d} 
\zeta(4+d) \Gamma(4+d) \; 
T^{4+d} \quad {\mbox{in the bulk,}}
\end{eqnarray}
which yields a ratio brane/bulk of
\begin{eqnarray}   
\frac{\Omega_{(3)}^2}{\Omega_{(3+d)}^2} \frac{\zeta(4)}{\zeta(4+d)} \frac{\Gamma(4)}{\Gamma(4+d)}  
\left(\frac{2\pi}{T R_H}\right)^{d}
\quad.
\end{eqnarray}
Inserting Eq. (\ref{tempD}) results in a ratio, dependent on $d$, which is of order $4 - 12$, for 
$d\leq 7$ with a 
maximum\footnote{This is due to the surface of the unit sphere having a maximum for $7$ dimensions.}  
for $d=4$. 
In addition
one has to take into account that there exists a much larger number of particles in the 
Standard Model than those which are allowed in the bulk. Counting up all leptons, 
quarks, gauge bosons, their anti particles and summing over various 
quantum numbers yields more than $60$. This has to be contrasted with only one graviton in the 
bulk\footnote{In the effective description on the brane, 
one might argue that there exists a large number of gravitons
with different masses. We have already taken this into account by integrating over the higher
dimensional momentum space.} which strongly indicates that most of the radiation goes into
the brane. 

However, the exact value will depend on the correct spin
statistics and in general, the ratio also
depends on the temperature of the black hole as can e.g. be seen in Figure \ref{mdot}, left. We also
see that for masses close to the new fundamental scale the evaporation rate is higher for a smaller
number of extra dimensions. This is due to the fact that in this region the evaporation rate is
better approximated by a higher dimensional version of Wien's law. 
For a precise calculation one also has to take into account that the presence of the gravitational
field will modify the radiation properties for higher angular momenta through backscattering at the
potential well. These energy dependent
greybody factors can be calculated by analyzing the wave equation in the higher dimensional spacetime
and the arising absorption coefficients. 
A very thorough description of these evaporation characteristics has been given in \cite{Kanti:2004nr}
which confirms the expectation that the bulk/brane evaporation rate is of comparable magnitude but
the brane modes dominate.

For recent reviews on TeV-scale black holes see also \cite{Landsberg:2002sa} and
references therein.

\subsection{Observables of Black Holes}

Due to the high energy captured in the black hole, the decay of such an object is a very spectacular
event with a distinct signature. The number of decay products, the multiplicity, is high compared to
Standard Model processes and the thermal properties of the black hole will yield a high sphericity
of the event. Furthermore, crossing the threshold for black hole production causes a sharp cut-off for
high energetic jets as those jets now end up as black holes instead, and are re-distributed
into thermal particles of lower energies. Thus, black holes will give a clear signal.

It is apparent that the consequences of black hole production are quite disastrous for the
future of collider physics. Once the collision energy crosses the threshold for black hole production, 
no further
information about the structure of matter at small scales can be extracted. As it was put by
Giddings and Thomas \cite{Giddings3}, this would be ''{\sl the end of short distance physics}''.
  
\vspace*{-1.5cm}

\begin{figure}[h]
\hspace*{-0.5cm}
\parbox{6.5cm}
{\vspace*{1cm}
\epsfig{figure=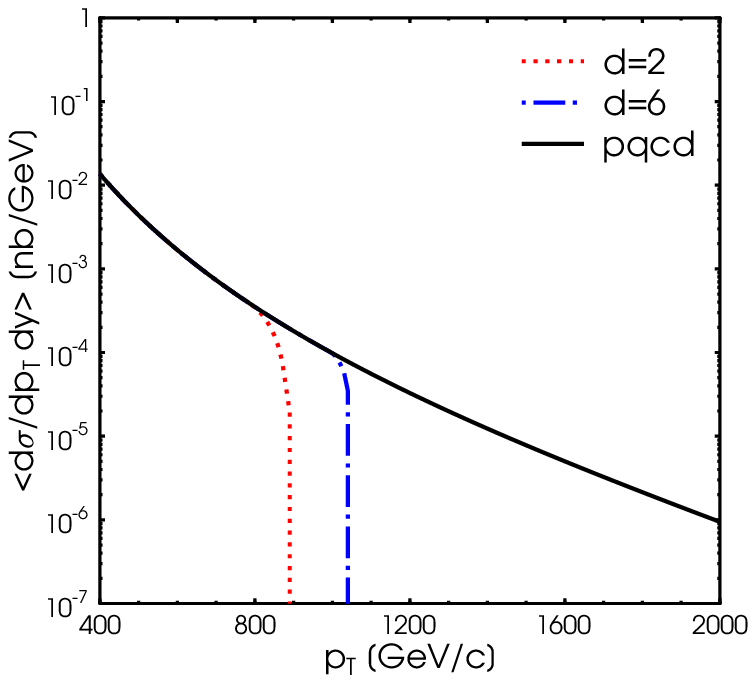, width=6.5cm}}

\vspace*{-6cm}
\hspace*{6cm}  
\parbox{8cm}
{\epsfig{figure=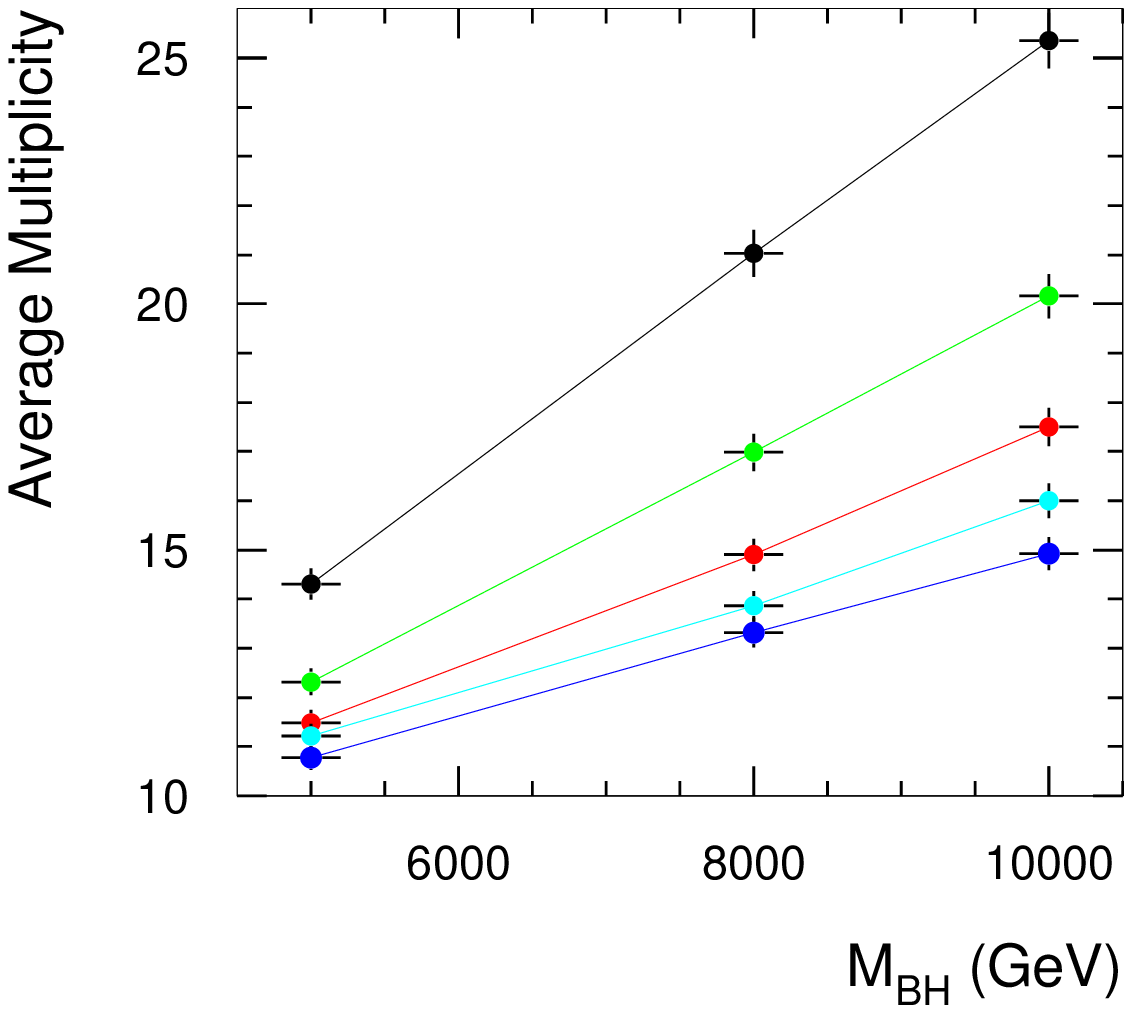, width=8cm}}

\centering
\parbox{12cm}
{\caption{Left: The onset of black hole production in high energy collisions results in a cut off 
for high $p_T$-jets. Here, it is $\sqrt{s}= 14$~TeV and $M_{\rm f}=1$~TeV. The average over the
rapidity  region $|y|\leq 1$ is taken. Right: This plot from \cite{BHex} shows the multiplicity of black hole
events for various numbers of extra dimensions $d =$ 2 (black), 3 (green), 4 (red), 
5 (cyan)  and 6 (blue).  
\label{ptbh}} }
 
\end{figure}


Figure \ref{ptbh}, left shows the expected cross section for jets at high transverse momentum with  
an onset of black hole production in comparison to the perturbative {\sc QCD} prediction. 
The cutoff is close by the new fundamental 
scale\footnote{We want to remind the reader that for qualitative arguments we have set the black hole
threshold to be equal to the fundamental scale.}  but
is hardly sensitive to the number of extra dimensions.

By now, several experimental groups include black holes into their search for physics beyond
the Standard Model.  
PYTHIA 6.2 \cite{PYTHIA} and the CHARYBDIS \cite{CHARYBDIS}
event generator allow for a simulation of black hole events and data reconstruction from the
decay products. Such analysis has been summarized in Ref. \cite{Atlas} and Ref. \cite{BHex}, respectively.
Ideally, the energy distribution of the decay products allows a determination of the temperature (by
fitting the energy spectrum to the predicted shape) as
well as of the total mass of the object (by summing up all energies).
This then allows to reconstruct the scale $M_{\rm f}$ and the number of extra dimensions. An example for
this is shown in Figure \ref{landsberg}, from \cite{dim}.

The quality of the determination, however, depends on the uncertainties in the
theoretical prediction as well as on the experimental limits e.g. background from
Standard Model processes. Besides the formfactors of black hole production and
the greybody factors of the evaporation, the largest theoretical uncertainties turn
out to be the final decay and the time variation of the temperature. In case the
black hole decays very fast, it can be questioned \cite{dim} whether it has time to readjust
its temperature at all or whether it essentially decays completely with its initial
temperature.

Also, the determination of the properties depends on the
number of emitted particles. The less particles, the more difficult the analysis. 

Figure \ref{ptbh} right, from \cite{BHex}, shows the number of expected decay products of a black hole
event. For larger $d$ with a fixed mass, the temperature increases which leads to less higher energetic
particles.  In Ref. \cite{BHex} it was also pointed out that the high sphericity even at 
low multiplicity allows clear distinction from Standard Model and/or SUSY events. In a
simulated test case, it was found that the fundamental scale can be 
determined up to 15\% and the number of extra dimensions up to $\pm$0.75.
 

\begin{parbox}[h]{14cm}

\begin{figure}[h]

\parbox{5.5cm}{\caption{This plot from \protect\cite{dim} shows an example for
a reconstruction of the parameters from simulated black hole events. Here, it was
assumed that the black hole decays so fast that the temperature can not be readjusted
and is therefore time independent.
\label{landsberg}}}

\vspace*{-3.8cm}
\hspace*{6.0cm}
\parbox{8.0cm}{
\epsfig{figure=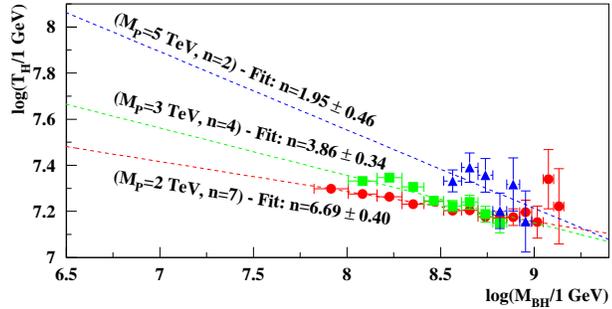,width=8.0cm}
 
} 
\end{figure}

\end{parbox}


\section{Black Hole Relics}

The final fate of these small black holes is difficult to estimate.
The last stages of the evaporation are closely connected to the  information loss puzzle.
The black hole emits thermal radiation, whose sole property is the
temperature, regardless of the initial state of the collapsing matter. 
So, if the black hole completely decays into statistically
distributed particles, unitarity can be violated. This happens when 
the initial state is a pure quantum state and then evolves into a mixed 
state \cite{Novi,Hawk82,Preskill}.

When one tries to avoid the information loss problem two possibilities are left. 
The information
is regained by some unknown mechanism or a stable black hole remnant is formed 
which keeps the
information. Besides the fact that it is unclear in which way the information should escape the horizon 
\cite{escape} there are several other arguments for 
black hole relics \cite{relics}:
\begin{itemize}
\item
The uncertainty relation: The Schwarzschild radius of a black hole with Planck mass 
is of the order  of the Planck length. Since the Planck length is the wavelength corresponding to a particle of
Planck mass, a problem arises when the mass of the black hole drops below Planck mass. 
Then one has trapped a higher mass, $M\geq M_{\rm f}$, inside a volume which is smaller than allowed by the uncertainty 
principle \cite{39}. To avoid this problem, Zel'dovich has proposed that black holes with masses below 
Planck mass should be associated with stable elementary particles \cite{40}. 

\item
Corrections to the Lagrangian: The introduction of additional terms, which are quadratic in the curvature, yields a dropping of the evaporation temperature towards zero \cite{Barrow,Whitt}. This holds also for 
extra dimensional scenarios \cite{my2} and is supported  by calculations in the low energy limit 
of string theory \cite{Callan,Stringy}.

\item
Further reasons for the existence of relics have been suggested to be black holes with axionic charge
\cite{axionic}, the modification of the Hawking temperature due to quantum hair \cite{hair} or magnetic 
monopoles \cite{magn}. Coupling of a dilaton field to gravity also yields relics, with detailed 
features depending on the dimension of space-time \cite{dilaton1,dilaton2}.

\end{itemize}

Of course these relics, which have also been termed Maximons, Friedmons, Cornucopions, Planckons or 
Informons\footnote{Though there are subtle but important differences between these objects.}, are not a miraculous remedy but bring some problems on their own. Such is e.g.
the necessity for an infinite number of states which allows the unbounded information
content inherited from the initial state.

However, let us consider a very simple model which will help us to understand the 
features which lead to the formation of a stable black hole remnant. 
We will suppose that the spectrum of the radiation emitted by the black hole is
discretized by geometrical means. The smaller the black hole is, the smaller is the
region of space time out of which the radiation emerges and we expect the finite
size of the black hole to reflect in the features of the radiation. The details
of this ansatz have been worked out in \cite{Hossenfelder:2003dy}.
 
In particular, for the derivation of the black hole's Hawking radiation, it
turns out that the black hole acts like a black body. These analogy to geometrical optics
can be put even further, see e.g. \cite{Chandra}. 

We split the solution for the $d+3$ dimensional wave equation of the black hole modes into an amplitude
proportional to $\propto 1/r^{d+1}$, an angular dependence and a radial part. We then assume
that the spectrum of the modes can only have discrete values due to the geometric boundary
conditions  $k_l = \pi l/R_H$, as depicted in Figure \ref{quant}.


\begin{parbox}[h]{14cm}

\begin{figure}[h]

\parbox{7.5cm}{
\caption{The black hole can not emit wave length that do not fit to the
horizon size. The spectrum of the radiation is geometrically quantized
in discrete steps. If the black hole shrinks, the lowest energetic
mode can already exceed the total energy and thus, can not
be emitted.
\label{quant}}
 }

\vspace*{-2.8cm}
\hspace*{8.0cm}
\parbox{6.0cm}{
\epsfig{figure=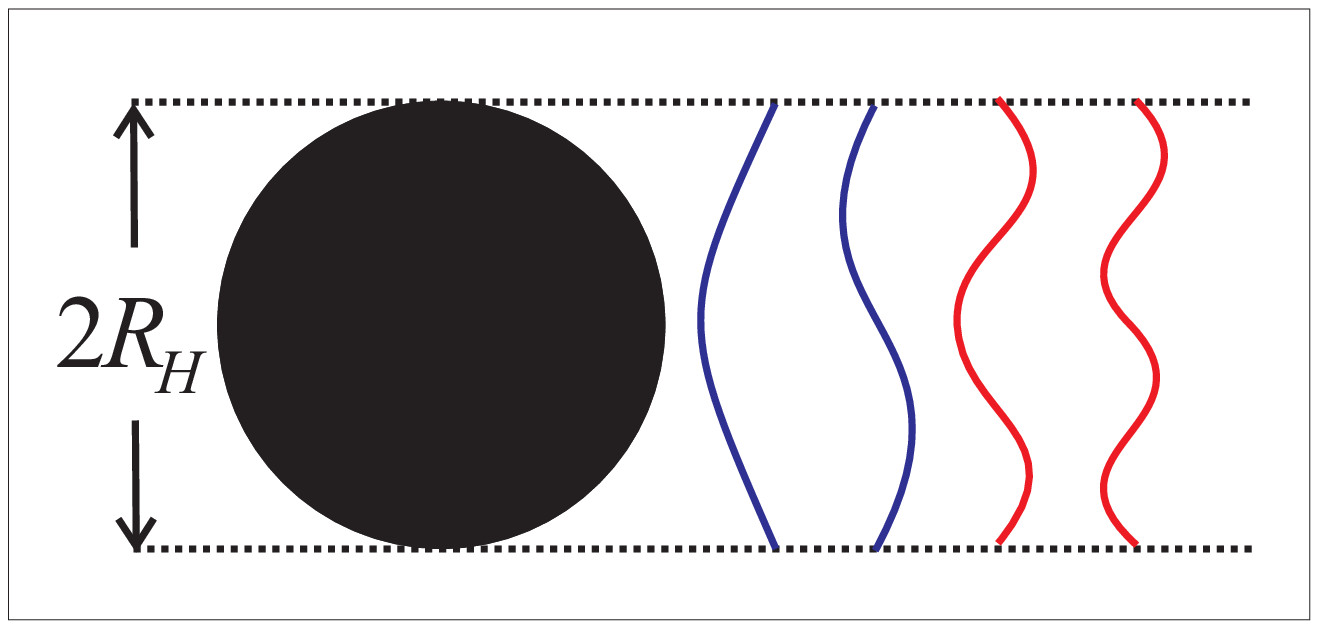,width=6.0cm}
 
}
\end{figure}

\end{parbox}

\begin{figure}[h]
\centering
\epsfig{figure=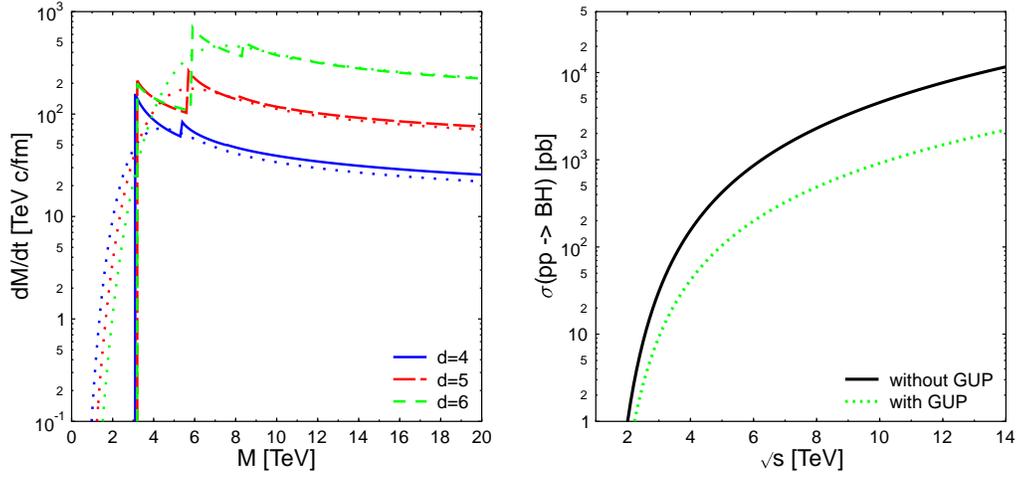,width=14cm}
\parbox{13cm}{\caption{The left plot\protect\cite{Hossenfelder:2003dy}
 shows the evaporation rate with discrete energy spectrum 
in comparison to the continuous case (dotted) for various $d$. The right plot, from \protect\cite{Hossenfelder:2004ze}, 
shows the total cross section
for black hole production as a function of the center of mass energy $\sqrt{s}$ with and without
the generalized uncertainty principle ({\sc GUP}). For the expected {\sc LHC}-energies the cross sections is
lowered by about a factor 5. This result does barely depend on the number of extra dimensions.
\label{dmdtq}}} 
\end{figure}

The particle spectrum Eq. (\ref{mehrteil}) is then replaced by a spectrum for the
discrete energy values $\omega_l =  l \Delta \omega $ with step size $\Delta \omega = \pi / R_H$
\begin{equation} \label{mehrteilq}
n(l) = 
\sum_{j=1}^{\lfloor \frac{M}{l \Delta \omega} \rfloor} \frac{{\rm exp}[S(M-j l \Delta \omega)]}{{\rm exp}[S(M)]} 
\Theta\left(M - l \Delta \omega \right) \quad.
\end{equation}
Here, the Heaviside-function $\Theta$ cuts off the sum when the energy of one particle ($j=1$) exceeds the
whole mass of the black hole. The energy density is then given by a summation instead of an integration
\begin{eqnarray} \label{epsq}
\varepsilon &=&  \frac{\Omega_{(d+3)}}{(2 \pi)^{3+d}}  
\Delta \omega  
\sum_{l=1}^{ \lfloor \frac{M}{\Delta \omega} \rfloor } n(l)  
\left( l \Delta \omega \right)^{d+3}  
\;\;,
\end{eqnarray}
which does as before result in an expression for the evaporation rate.
For large masses, $M \gg M_{\rm f}$, the sum becomes an integral and 
Eq.(\ref{epsq}) reduces to the continuous case.   
Figure \ref{dmdtq}, left, shows the evaporation rate with geometrical quantization in contrast to
the continuous result (compare to Figure  \ref{mdot}).


With increasing $M$, the steps between the energy levels become smaller.
If it is possible to occupy an additional level,  the evaporation rate has a step.
In the mass region of interest for us,  it is $R_H \approx 1/M_{\rm f}$ and thus the steps
are  $\approx \pi$TeV. 
 
The lowest energetic mode possible corresponds to the largest wavelength which
fits into the black holes horizon. The smaller the black hole is, the smaller this
wavelength has to be and thus, the energy of the lowest level increases. At black
hole masses close to the Planck mass, it happens that even the energy of the
lowest level would exceed the total energy of the black hole and can not be emitted.
 
The evaporation therefore will proceed in quantized steps and stops at a finite
mass value. A relic is left.

The production of a relic instead of a final decay of the black hole would
increase the missing $p_T$ in the particle collision. Also, a certain fraction of these
objects will be charged which allows for a direct detection.

\section{The Minimal Length}
\label{MinimalLength}

Gravity itself is inconsistent with physics
at very short scales.
The introduction of gravity into into quantum field theory appears to spoil their renormalizability and 
leads to incurable divergences. It has therefore been suggested that gravity
should lead to an effective cutoff in the ultraviolet, i.e. to a minimal observable length.
It is amazing enough that all attempts 
towards a fundamental theory of everything necessarily seem to imply the existence 
of such a minimal length scale. 

The occurrence of the minimal length scale has to be expected from very general reasons 
in all theories at high energies which attempt to include effects of
quantum gravity. This can be understood by a phenomenological argument which has been examined in various ways. 
Test particles of a sufficiently high energy to resolve a distance as small as the Planck 
length, $l_{\rm p}=1/m_{\rm p}$, 
are predicted to gravitationally curve and thereby to significantly disturb the very spacetime structure
which they are meant to probe. Thus, in addition to the expected quantum uncertainty, there is another
uncertainty caused which arises from spacetime fluctuations at the Planck scale.

Consider a test particle, described by a wave packet with a mean Compton-wavelength {$\lambda$}. 
Even in standard quantum mechanics, the particle suffers from an uncertainty in position $\Delta x$ 
and momentum $\Delta p$, given by the standard Heisenberg-uncertainty relation  $\Delta x \Delta p \geq 1/2$. 
Usually, every sample under investigation can be resolved by using beams of  
high enough momentum to focus the wave packet to a width below the size of the probe, $\Delta x$, which
we wish to resolve. 
The smaller the sample, the higher the energy must become and thus, the 
bigger the collider.

General Relativity tells us that a particle with an energy-momentum $p \sim 1/\lambda$ in a
volume of spacetime $\Delta x^3$ causes a fluctuation of the metric $g$. Einstein's Field
Equations (\ref{EFG}) allow us to relate the second derivative of the metric to the energy density
and so we can estimate the perturbation with
\begin{eqnarray}
\frac{\delta g}{\Delta x^2} \sim l_{\rm p}^2 \frac{p}{\Delta x^3} \quad.
\end{eqnarray}
This leads to an additional fluctuation, $\Delta x'$, of the spacetime coordinate frame of order 
$\Delta x' = \delta g \Delta x \approx l_{\rm p}^2 p$, which increases with momentum and becomes
non-negligible at Planckian energies. 

But from this expression we see not only that the fluctuations can no longer be neglected at
Planckian energies and the uncertainty of the measurement is amplified. In addition,
we see that a focussation of energy of the amount necessary to resolve the Planck scale
leads to the formation of a black hole whose horizon is located at $\delta g \approx 1$, should
the matter be located inside. Since the black hole radius has the property to expand linearly
with the energy inside the horizon, both arguments lead to the same conclusion: a minimal
uncertainty in measurement can not be erased by using test particles of higher energies. 
It is always  $\Delta x > l_{\rm p}$: at low energies because the Compton-wavelengths
are too big for a high resolution, at high energies because of strong curvature
effects.   

Thus, to first order the uncertainty principle 
is generalized to
\begin{eqnarray}
\Delta x \gtrsim \frac{1}{\Delta p} + \mbox{const.}~~ l_{\rm p}^2 {p}\quad. \label{gup}
\end{eqnarray}
Without doubt, the Planck scales marks a thresholds beyond which the old description of space-time 
breaks down and new phenomena have to appear.

More stringent motivations for the occurrence of a minimal length are manifold. 
A minimal length can be found in String Theory, Loop Quantum Gravity and Non-Commutative Geometries.
It can be derived from various studies of thought-experiments, from black hole physics, the holographic
principle and further more. Perhaps the most convincing argument, however, is that there seems to
be no self-consistent way to avoid the occurrence of a minimal length scale. For reviews see e.g. 
\cite{Garay:1994en}.
 
Instead of finding evidence for the minimal scale as has been done in numerous studies, on can 
use its existence as a postulate and derive extensions to quantum theories 
with the purpose to examine the arising properties in an effective model.

Such approaches have been undergone and the analytical properties of the resulting theories have been 
investigated closely \cite{Kempf:1994su}. 
In the scenario without extra dimensions, the derived modifications are 
important mainly for structure formation
and the early universe \cite{earlyuniverse}. 

The importance to deal with a finite resolution
of spacetime, however, is sensibly enhanced if we consider a spacetime with 
large extra dimensions. In this case, the new fundamental mass scale is lowered and thus, the minimal length
is raised. The effects of a finite resolution of spacetime should thus become important in the
same energy range in which the first effects of gravity are expected.
Some consequences of the inclusion of the minimal length into the model of extra dimensions
have been worked out in \cite{Bhattacharyya:2004dy}.


\begin{parbox}[h]{15cm}

\begin{figure}[h]

\parbox{7cm}{
A generalized uncertainty principle, as follows from the existence of a
minimal length scale, makes it increasingly difficult to focus the wave functions of the
colliding particles. This is sketched in Figure \ref{bhgup}. It requires higher energies to 
achieve an energy density inside the Schwarzschild radius which is sufficient for a collapse 
and the creation of a black hole. 

The consequences of the generalized uncertainty 
for the total black hole cross section and the number of produced black holes
have been derived in \cite{Hossenfelder:2004ze}
and are depicted in Figure \ref{dmdtq}, right. As we expect, the cross section 
is lowered. For the estimated {\sc LHC}-energies this results in a suppression of the
black hole production by a factor $\approx 5$.
 }

\vspace*{-8cm}
\hspace*{7.5cm}
\parbox{6.5cm}{
\epsfig{figure=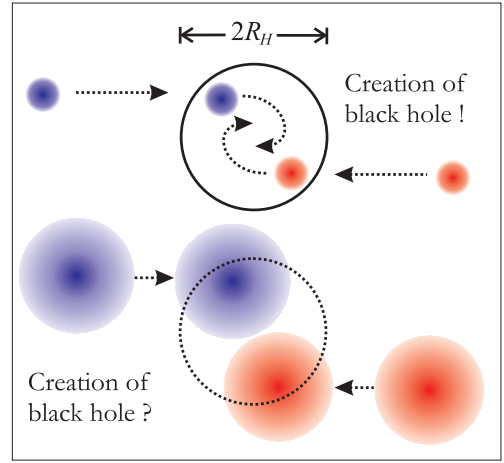,width=6.5cm}
\caption{Schematic figure for parton scattering with and without generalized uncertainty
\label{bhgup}.}
}
\end{figure}

\end{parbox}


The arising modifications at high energies do also influence the production of black 
holes. 
Using the generalized uncertainty principle, 
the properties \cite{Cavaglia:2004jw} of the evaporation process are modified when
the black hole radius approaches the minimal length
which decreases the number of decay particles and leads to the formation of
a stable remnant. 

\section{Are Those Black Holes dangerous?}
\label{Dangerous}

''{\sl Big Bang Machine: 
Will it destroy Earth?, 
Creation of a black hole on Long Island?

\noindent
The committee will also consider [...] that the colliding 
particles could achieve such a high density that they would 
form a mini black hole. In space, black holes are believed 
to generate intense gravitational fields that suck in all surrounding matter. 
The creation of one on Earth could be disastrous. [...]}''
\begin{flushright}The London Times, July 18th, 1999   
\end{flushright}

Considered the fact that black holes are known to swallow whole galaxy centers, the
question might be raised whether the tiny black holes, produced in the laboratory,
would also accrete mass, swallow it and grow until everything in reach would be 
vanished beyond the event horizon.

Fortunately, this question can without doubt be answered with 'No'. These earth
produced black
holes still have a very small gravitational attraction. An energy of some TeV might
be huge for collider physics but for gravitational purposes it is tiny.
The strength of the gravitational force, acting on electrons orbiting around a
nucleus, is even in the presence of extra dimensions still by a factor $\approx 10^{20}$ 
weaker than the electromagnetic force\footnote{The precise value depends on the
number of extra dimensions.}.

We can estimate the possibility of black holes gaining mass based on our earlier given
results. Looking at Figure \ref{mdot}, we see that the mass loss ratio, ${\rm d} M_-/{\rm d} t$
 of the black hole is at least $\approx 1$~TeV $c$/fm\footnote{1 fm/$c$ is $\approx 10^{-23}$~sec and is the typical
timescale for high energy collisions.}, for a large number of extra
dimensions it would be even larger. As previously stated, the gravitational attraction
of the black hole is negligible. Nevertheless, is the black hole placed in a medium which
is considered to be extremely dense and hot in earthly terms, it will gain mass by particles
crossing the horizon simply through thermal motion. This mass gain ratio can be estimated
to be
\begin{eqnarray}
\frac{dM_+}{dt} \sim \pi R_H^2 T_{\rm m}^4 \quad, 
\end{eqnarray}
where $T_{\rm m}$ is the temperature of the medium and the fore-factor is the surface of the black hole
as seen on our submanifold\footnote{Note, that is not the total surface of the black hole as this is
a $d+2$ dimensional hypersurface.}. Inserting $\rho=T_{\rm m}^4$ to be the most dense medium produced
on earth, e.g. a Quark Gluon Plasma, we have typically $\rho \sim 100$~MeV/fm$^3$. Using as a characteristic
value $R_H = 1/M_{\rm f} \sim 1/$TeV, we find that
\begin{eqnarray} \label{massgain}
\frac{{\rm d}M_+}{{\rm d}t} \sim 10^{-9} \frac{{\rm GeV}c}{\rm fm} \ll \frac{{\rm d}M_-}{{\rm d}t} \quad, 
\end{eqnarray}
and so the black hole will loose mass much faster than it can gain mass.

Let us consider a further extreme case: a black hole, created by an ultra high high energetic
cosmic ray, which traverses a neutron star. Cosmic ray events can occur at energies as
high as $10^8$~TeV and thus, can yield black holes with an
ultra high $\gamma$-factor, $\gamma \sim E/M \sim 10^7$. This means, 
the black hole has an ultra high speed and due to the effects of 
Lorentz-contraction, it will see the medium of the neutron star in an even higher density.
Or, to put it different, in the reference frame of the neutron star the black hole travels fast, undergoes
the relativistic time dilation and evaporates slower.

In the reference frame of the black hole, the mass loss is still given by the above estimation 
but it now sees a medium with density $\gamma^2 \rho$.
The mass gain during passage of the neutron star is then increased by a factor $\gamma^2 \sim 10^{14}$.
By multiplying Eq. (\ref{massgain}) with this factor we see that in this case, the mass gain become 
comparable to the mass loss! The black hole thus has a chance to eat up a significant part of the
neutron star. 
This process has been investigated in \cite{Neutronstar}. 
However, the estimation we have given
here does only apply immediately after the creation of the black hole. Once it starts growing, it
will become more massive and slow down. This in turn will decrease its $\gamma$-factor, the mass loss again 
becomes dominant and the black hole decays. A continuing growth is possible only in a very constrained parameter
range which is already excluded \cite{Neutronstar}. 
\pagebreak 

\noindent 
\color{red}
\framebox{
\color{black}
\begin{parbox}{14cm}  
{ {\sc \large Prejudices against Black Holes}
\bigskip
 
{\underline{\sc Black holes are mathematical trash:}}
Wrong! The existence of an event horizon is a perfectly well defined space time property.
Moreover, their formation follows from such general assumptions that to present 
knowledge there is no way to avoid them. General Relativity {\sl requires} black holes. 
Since the horizon radius is proportional to the mass $M$, it follows that the
density needed for the horizon formation is proportional to $M/M^3$. This means, for large masses, the
density can be arbitrarily small.

\bigskip

{\underline{\sc Black holes are speculative:}} Wrong! The evidence for a
black hole in the center of our galaxy (Sag. A$^*$) is overwhelming \cite{SagA}. More than 20 other
candidates in our galaxy \cite{McClintock:2003gx} and 10 further 
super massive black holes \cite{Kormendy:2001hb}
have been established so far. 
The analysis of the
motion of the surrounding objects is used to determine the amount 
mass concentrated in their vicinity. The mass inside the region is so high that the only 
plausible explanation is a black hole and {\sl everything else} is highly
speculative \cite{McClintock:2004ji}.

\bigskip

{\underline{\sc Black holes are black:}} Wrong! The strong curvature effects in the vicinity of a black
hole can rip apart virtual particle pairs and result in particle emission in the Hawking-radiation process. 
For objects of astrophysical sizes, this temperature is too low to be observable. These large
black holes in general have rotating accretion disks surrounding them, in which the fast moving matter gets 
extremely hot and emits radiation.  

\bigskip

{\underline{\sc Black holes on earth are dangerous:}} Wrong! The black holes which could possibly be created on
earth would still have a very small mass and, connected with it, a very small gravitational
attraction. Their interaction with other particles is so weak that they can not grow, even
under extreme conditions, like inside a quark gluon plasma. In all cases, the evaporation
processes will dominate, causing the black hole to decay instead to grow.}
\end{parbox}
 }
 \color{black}

\section{Summary: What Black Holes Can Teach Us}
\label{Summary}

Black holes are a topic as fascinating as challenging. The investigations during the last century have led to
important insights. Thermodynamics of black holes taught us of a deep connection between
quantum theory, General Relativity and thermodynamics which has recently been studied   
further in the string/black hole correspondence and is a matter of ongoing research. 
Black holes and the puzzle of information loss have tought us about the 
possibility of stable black hole relics and the natural inclusion of a minimal length
scale.

The presence of additional compactified dimensions would allow
the production of tiny black holes in particle collisions.
Their investigation in experiments on earth would provide a direct
possibility to address the question of extra dimensions and the new fundamental
scale. But moreover, it would allow us to probe the phenomenology of physics at the
Planck scale and increase our knowledge about the underlying, yet to be found theory. Black holes can
teach us to find the missing link between general relativity and quantum theory. 

\section*{Acknowledgments}
I want to thank Marcus Bleicher, Stefan Hofmann and Horst St\"ocker for the inspiring 
discussions at Frankfurt University. I also thank Claudia Schumacher for critical remarks. 
I apologize for every missing citation on the topic.
This work was supported by the German Academic Exchange Service 
({\sc DAAD}), NSF PHY/0301998 and the {\sc DFG}. 

\begin{appendix}
\section{Useful Fomulae}

\begin{eqnarray}
\Gamma ^{\alpha}_{\; \; \; \mu
\nu} &=& \frac{1}{2} g^{\alpha \beta} (
\partial_\mu g_{\beta \nu} + 
\partial_\nu g_{\beta \mu} - 
\partial_\nu g_{\mu \nu} ) \label{Christoffel}\\
R^{\mu}_{\; \; \; \nu
\alpha \beta} &=& \partial_\alpha \Gamma ^{\mu}_{\; \; \; \nu
\beta}  - \partial_\alpha \Gamma ^{\mu}_{\; \; \; \nu
\alpha} + \Gamma ^{\mu}_{\; \; \; \rho
\alpha} \Gamma ^{\rho}_{\; \; \; \nu
\beta} - \Gamma ^{\mu}_{\; \; \; \rho
\beta} \Gamma ^{\rho}_{\; \; \; \nu
\alpha} \label{R1}\\
R^{\mu}_{\; \; \; \nu} &=& R^{\alpha}_{\; \; \; \mu
\alpha \nu} \label{R2}\\
R &=& g^{\mu \nu} R_{\mu \nu} \label{R3}
\end{eqnarray}

\section{Useful Constants}

{\small
\begin{table}[h]
\centering
\begin{tabular}{|l|c|cl|cl|}
\hline
Planck length & $l_{\rm p}$ & $\approx$ & $10^{-35}$~m & $\approx$ & $10^{-20}$~fm   \\ \hline 
Planck mass   & $m_{\rm p}$ & $\approx$ & $10^{-5}$~g  & $\approx$ & $1.2 \times 10^{16}$~TeV  \\ \hline 
Planck time   & $t_{\rm p}$ & $\approx$ & $10^{-43}$~s & $\approx$ & $10^{-20}$~fm/$c$  \\ \hline 
Mass of the sun   & $M_\odot$   & $\approx$ & $2\times 10^{33}$~g & $\approx$ & $2.4 \times 10^{54}$~TeV \\ \hline 
Radius of the sun & $R_\odot$   & $\approx$ & $7\times 10^{8}$~m  &          & \\ \hline 
Schwarzschild &&&&& \\  
radius of the Sun        & $R_{H}$ & $\approx$ & $3 \times 10^{3}$~m & &  \\ \hline 
c.o.m. energy  &&&&& \\ 
at the LHC & $\sqrt{s}$ & && $\approx$ & 14~TeV  \\ \hline 
\end{tabular}
\end{table} }

\end{appendix}

\small{
}  
\end{document}